%% file: y3-wl_image_sims.tex
\documentclass[fleqn,usenatbib]{mnras}

\usepackage{newtxtext,newtxmath}

\usepackage[T1]{fontenc}
\usepackage{ae,aecompl}

\usepackage{graphicx}	
\usepackage{amsmath}	
\usepackage{listings}
\usepackage[dvipsnames]{xcolor}
\usepackage{siunitx} 
\usepackage{bm}

\usepackage{eso-pic}

\AddToShipoutPictureBG*{%
  \AtPageUpperLeft{%
    \hspace{0.75\paperwidth}%
    \raisebox{-3.5\baselineskip}{%
      \makebox[0pt][l]{\textnormal{DES-2020-0550}}
}}}%

\AddToShipoutPictureBG*{%
  \AtPageUpperLeft{%
    \hspace{0.75\paperwidth}%
    \raisebox{-4.5\baselineskip}{%
      \makebox[0pt][l]{\textnormal{FERMILAB-PUB-20-628-AE}}
}}}%

\newcommand{\snr}{\ensuremath{S/N}}

\newcommand{\gobs}{\ensuremath{\bar{g}^{\mathrm{obs}}}}
\newcommand{\gobsa}{\ensuremath{\bar{g}^{\mathrm{obs}}_{\alpha}}}
\newcommand{\gtrue}{\ensuremath{g^{\mathrm{true}}}}
\newcommand{\gtruez}{\ensuremath{g^{\mathrm{true}}(z)}}
\newcommand{\gtruea}{\ensuremath{g^{\mathrm{true}}_{\alpha}}}
\newcommand{\gconst}{\ensuremath{g^{\mathrm{const}}}}

\newcommand{\fiducial}{\texttt{fiducial}}
\newcommand{\grid}{\texttt{grid}}
\newcommand{\gridtruedet}{\texttt{grid-truedet}}

\newcommand{\neffz}{\ensuremath{n_{\gamma}(z)}}
\newcommand{\neffiz}{\ensuremath{n_{\gamma,i}(z)}}
\newcommand{\neff}{\ensuremath{n_{\gamma}}}

\newcommand{\neffzmcal}{\ensuremath{n_{\gamma}^{\text{mcal}}(z)}}
\newcommand{\neffizmcal}{\ensuremath{n_{\gamma,i}^{\text{mcal}}(z)}}
\newcommand{\neffizmodel}{\ensuremath{n_{\gamma,i}^{\text{model}}(z)}}

\newcommand{\Nga}{\ensuremath{N_{\gamma}^{\alpha}}}
\newcommand{\Ngia}{\ensuremath{N_{\gamma,i}^\alpha}}
\newcommand{\Ngiamodel}{\ensuremath{N_{\gamma,i}^{\alpha,\text{model}}}}
\newcommand{\Ngamcal}{\ensuremath{N_{\gamma}^{\alpha,\text{mcal}}}}
\newcommand{\Ngiamcal}{\ensuremath{N_{\gamma,i}^{\alpha,\text{mcal}}}}

\newcommand{\gbarobs}{\ensuremath{\bar{\gamma}^{\text{obs}}}}
\newcommand{\gbarobsi}{\ensuremath{\bar{\gamma}^{\text{obs}}_i}}

\newcommand{\dz}{\ensuremath{\delta \bar{z}}}
\newcommand{\Dz}{\ensuremath{\Delta \bar{z}}}

\newcommand{\dx}[1]{\mathrm{d}{#1}\,}

\newcommand\eqn[1]{equation~\ref{#1}}

\newcommand\Eqn[1]{Equation~\ref{#1}}   

\newcommand\fig[1]{Figure~\ref{#1}}

\newcommand\sect[1]{Section~\ref{#1}}
\newcommand\tab[1]{Table~\ref{#1}}

\DeclareSIUnit \h {\mbox{$h$}}
\DeclareSIUnit \hinv {\mbox{$h^{-1}$}}
\DeclareSIUnit \deg {\mbox{deg}}
\DeclareSIUnit \msun {\mbox{$M_{\odot}$}}

\DeclareSIUnit \parsec {pc}
\DeclareSIUnit \kilaparsec {kpc}
\DeclareSIUnit \megaparsec {Mpc}
\DeclareSIUnit \gigaparsec {Gpc}
\DeclareSIUnit \arcmin {arcminutes}

\newcommand\sex{\textsc{SExtractor}}

\newcommand\mcal{\textsc{Metacalibration}}

\newcommand\photoz{photo-$z$}

\def\shrug{\texttt{\raisebox{0.75em}{\char`\_}\char`\\\char`\_\kern-0.5ex(\kern-0.25ex\raisebox{0.25ex}{\rotatebox{45}{\raisebox{-.75ex}"\kern-1.5ex\rotatebox{-90})}}\kern-0.5ex)\kern-0.5ex\char`\_/\raisebox{0.75em}{\char`\_}}}

\title[Blending shear and redshift biases]{Dark Energy Survey Y3 results: Blending shear and redshift biases in image simulations} 

\input{authors.tex}

\date{Accepted XXX. Received YYY; in original form ZZZ}

\pubyear{2020}

\begin{document}
\label{firstpage}
\pagerange{\pageref{firstpage}--\pageref{lastpage}}
\maketitle

\begin{abstract}
As the statistical power of  galaxy weak lensing reaches percent level precision, large, realistic and robust simulations are required to calibrate observational systematics, especially given the increased importance of object blending as survey depths increase. To capture the coupled effects of blending in both shear and photometric redshift calibration, 
we define the effective redshift distribution for lensing, \neffz, and describe how to estimate it using image simulations.
We use an extensive suite of tailored image simulations to characterize the performance of the shear estimation pipeline applied to the Dark Energy Survey (DES) Year 3 dataset. We describe the multi-band, multi-epoch simulations, and demonstrate their high level of realism through comparisons to the real DES data. We isolate the effects that generate shear calibration biases by running variations on our fiducial simulation, and find that blending-related  effects are the dominant contribution to the mean multiplicative bias of approximately $-2\%$.  By generating simulations with  input shear signals that vary with redshift, we calibrate biases in our estimation of the effective redshfit distribution, and demonstrate the importance of this approach when blending is present. We provide corrected effective redshift distributions that incorporate  statistical and systematic uncertainties, ready for use in DES Year 3 weak lensing analyses. 
\end{abstract}
\vspace{1cm}


\makeatletter
\def \blfootnote{\xdef\@thefnmark{}\@footnotetext}
\makeatother

\blfootnote{$^{\dag}$ E-mail: nm746@cam.ac.uk}
\blfootnote{ Affiliations are listed at the end of the paper.}



\section{Introduction}\label{sec:intro}

While weak gravitational lensing of galaxies has enormous potential as a cosmological probe (e.g. \citealt{weinberg13,albrecht2006}),  measurements of the weak lensing shear have proven to be extremely difficult in practice (e.g. \citealt{mandelbaum14}). The shear manifests as a small distortion in the observed shape of a galaxy. Its measurement is subject to numerous biases, and requires accurate calibrations of many properties of the input images. Typically, these biases have been quantified by assuming a linear relation between component $x$ of the measured shear $\gobs_x$ (averaged over an ensemble of galaxies), and component $y$ of the true shear $\gtrue_y$ (e.g. \citealt{heymans06})
\begin{equation}
\gobs_x = (1+m_{xy})\gtrue_y + c_x,
\end{equation} 
where $m_{xy}$ is known as the multiplicative bias, and $c$ is known as the additive bias. This linear relation is expected to hold in the weak lensing regime, where $\gtrue_y$ is small (and so contributions of order $(g_y^{\mathrm{true}})^2$ can be neglected). One can also consider the multiplicative term as quantifying the linear response, $R_{xy} \equiv 1+m_{xy}$ of the shear estimate to a change in the input shear i.e. 
\begin{equation}
R_{xy} \equiv \frac{\partial \gobs_x}{\partial \gtrue_y}.
\end{equation}
The off-diagonal elements of $m_{xy}$ (and $R_{xy}$) are often assumed (and empirically found) to be zero, allowing for the more common notation where $m_x \equiv m_{xx}$.
In the following we will usually drop the shear component subscripts for brevity, with expressions involving \gobs\ and \gtrue\ generally holding for either component of the shear.  

The most thoroughly studied biases in weak lensing measurements have mainly been performed with simulations of isolated objects. These include noise bias (e.g. \citealt{refregier12, kacprzak12}, model bias (e.g. \citealt{voigt10}), selection biases (e.g. \citealt{Kaiser2000,bernstein02,hirate03}), and biases from miscorrecting for the image point-spread function (PSF) (e.g. \citealt{paulinhenriksson08}). These problems were tackled by community-driven efforts like the STEP \citep{heymans06, step2} and GREAT \citep{great08,great10,great3} challenges, and aided by the development of the widely used \textsc{GalSim}\footnote{\url{https://github.com/GalSim-developers/GalSim}}  software for simulation of astronomical images \citep{rowe14}. For isolated objects, the aforementioned biases  have largely been solved by methods like \mcal\ \citep{huff17, sheldon17} and ``Bayesian Fourier Domain'' (BFD) \citep{bernstein2016}, at least for sufficiently well understood data (i.e. with accurately-characterized noise and background levels and PSF). The \mcal\ method is particularly powerful because it does not rely on calibration simulations, which inevitably rely on assumptions about the  properties of the faint, often poorly resolved galaxies used in weak lensing analyses. 

More recently, some studies have begun to study shear calibration biases in the context of multiple objects and blending. It has generally been assumed that in this case the use of image simulations will be essential, and these have been used for the calibration of recent weak lensing cosmology analyses, for example by \citet{fenechconti2017,kannawadi2019} (for the Kilo-Degree Survey\footnote{\url{http://kids.strw.leidenuniv.nl/index.php}}), \citet{mandelbaum2018} (for the Hyper Suprime-Cam Subaru Strategic Program\footnote{\url{https://hsc.mtk.nao.ac.jp/ssp/}}), and \citet{samuroff2018, kacprzak2020} (for DES Year 1 analyses). Works such as   \citet{hoekstra2017} and \citet{martinet19}, with an eye to deeper upcoming datasets, have used image simulations to study effects such as the impact of undetected galaxies on the shear calibration. 

In parallel to these simulation-based calibration efforts,  \citealt{sheldon2020} developed new measurement methodology,  \textsc{metadetection}, which corrects for much of the impact of  blending, in particular the significant shear biases imparted by detection and deblending algorithms, as well as the impact of blending at the shape measurement stage. The \textsc{metadetection} method does not require simulation-based calibration, and exhibits extremely low levels of shear calibration bias even on (constant shear) simulations designed to match the depth of Rubin Observatory Legacy Survey of Space and Time\footnote{\url{https://www.lsst.org/}} (LSST) data. 

\begin{figure*}
\includegraphics[width=1.5\columnwidth]{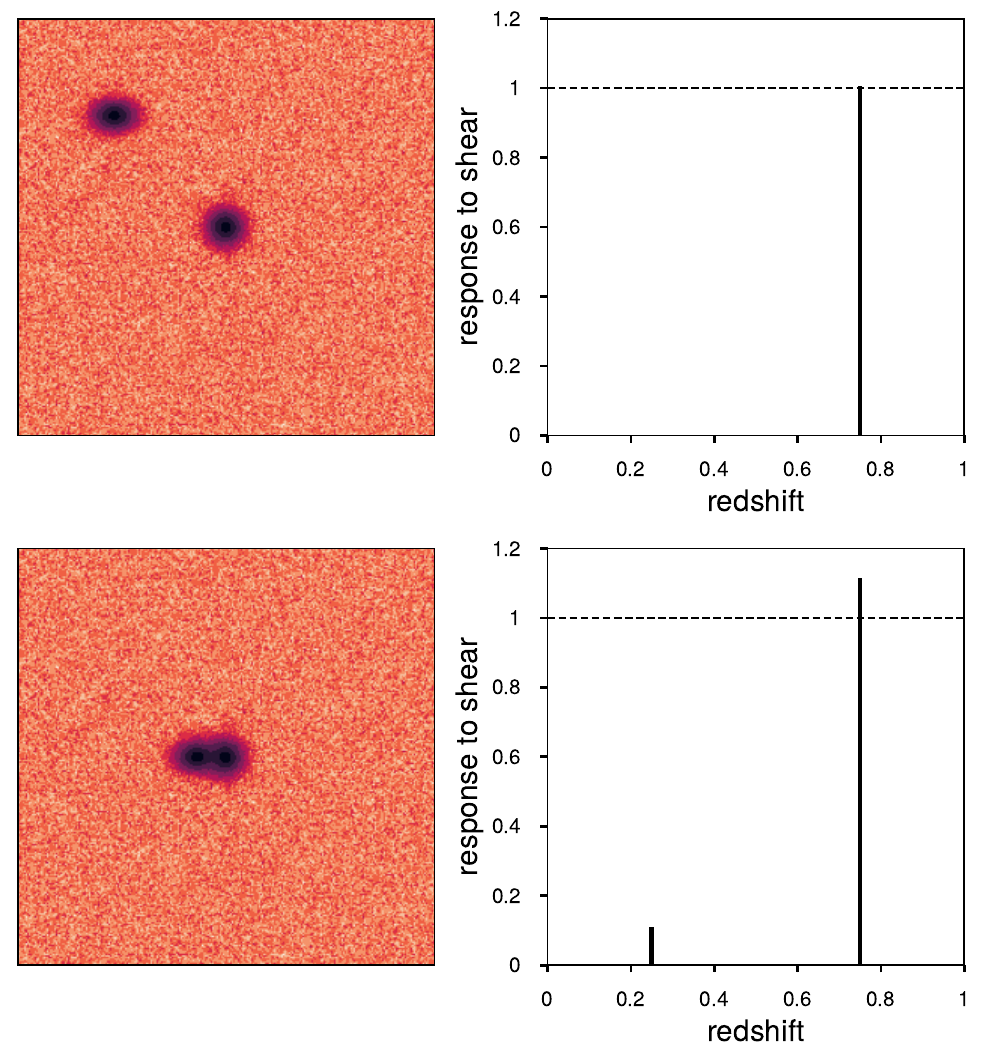}
\caption[]{
  Simple example of the interplay between blending, shear calibration, and photometric redshift distributions. In both rows, the left panel shows an image of a pair of simulated galaxies, with the central object at higher redshift, $z = 0.75$, and a neighboring object at low redshift, $z=0.25$. The right-hand panels show the response of the shape measurement of the central ($z=0.75$) object, to applied shear, as a function of the redshift at which that shear is applied. In the top-row, where the objects are not blended, the the high-redshift object only responds to a shear at its own redshift, and since we use \mcal\ shear estimation, the response is unity (i.e. the shear estimation is unbiased) to very high precision.
  In the bottom right panel, we show the response when the objects are blended. We see two effects here. First, due to blending, the \mcal\ estimator is now biased (the right peak at z=0.75 has non-unit height). Second, the high-redshift object now responds to input shears at other redshifts (the small peak at z = 0.25). We show in this work that these responses define the effective redshift distribution for lensing predictions, in addition to quantifying the multiplicative bias of the measurements.}
\label{fig:simplemod}
\end{figure*}

Current and future surveys will have large amounts of blending of objects at different redshifts (e.g. \citealt{dawson2016}). The component galaxies in blended systems will therefore often experience different shears. As we will discuss in \sect{sec:shear_biases}, the impact of this on weak lensing statistics cannot be fully accounted for by the use of simplified constant shear simulations used thus far in the field to calibrate shear measurements, or  corrections from shear estimation methods like \textsc{metadetection}. In order to gain intuition into the possible effects, consider the following simplified situation, shown in \fig{fig:simplemod}. In both panels we input a pair of   galaxies at different redshifts; for the sake of illustration we arbitrarily set one at low redshift ($z = 0.25$) and one at high redshift ($z = 0.75$). The high redshift galaxy is placed at the center of the stamp in both panels. For simplicity, we have fixed both galaxies to be round before lensing. In the top panels, the galaxies are not blended together, and in the bottom panels they are. In both cases we assume we can unambiguously detect two separate objects and precisely know their centroids. 

Let us consider the response to shear of the central ($z=0.75$) object in each stamp. We can apply shear separately at the two redshifts from which the light in the image is sourced (which in this case just means applying shear separately to the two galaxies present in the stamp). The right-hand panels of \fig{fig:simplemod} show the response to shear of the measured shape of the central galaxy, as a function of the redshift of the applied shear. The response is defined here as above, as $R\,{=}\,{\scriptstyle\left(\partial\gobs/\partial\gtrue\right)}$. We estimated these responses numerically from our simple simulations using the \mcal\ method.  In the top right panel, we see that the high redshift object does not respond to the shear of the low redshift one (it is zero at $z = 0.25$) and has a unit response to a shear applied at its own redshift (the peak at $z = 0.75$). This result makes sense since \mcal\ is known to be unbiased at high precision for idealized cases such as this, and the objects do not overlap.

The more interesting case is when the galaxies are blended, shown in the lower  panels of \fig{fig:simplemod}. In this case we see that the high redshift object responds to the shear of the low redshift object (the small peak at $z = 0.25$). It also has an apparent greater than unity response to the shear applied at its own redshift (the peak at $z = 0.75$ that is greater than one). Both of these effects are due to the galaxy being blended with the low redshift neighbor. The latter effect is likely due to the positioning of the neighboring galaxy in the positive $g_1$ direction, which is the shear component for which we compute the response. 

However, the  response of the high redshift object to a low redshift shear is a qualitatively different effect, distinct from standard multiplicative biases due to blending or detection; it is a bias that depends not only on the presence of the neighbor, but also on the shear applied to the neighbor. This indicates that the shape measurement of the high redshift object is carrying information about the low redshift shear.

In fact, we assert that it is the response to shear that defines how we should weight the redshifts to which we assign the shear information 
for a given object in a weak lensing analysis. This insight is a key subject of this work, where we will definitively
measure these effects in simulations of the DES Year 3 (Y3) analysis (note that ``Year 3'' includes the first three years of DES observations). 
There are also important implications for analysis of future surveys. As the amount of blending increases with increased depth, 
inferring the redshift distribution relevant for lensing and the shear calibration biases will be a joint
analysis task. In the example above, we have described how a single detected object can have a (non-unity) response to shear at multiple redshifts. This effect cannot be fully described by the traditional multiplicative bias, $m$. The shear calibration and the effective redshift distribution cannot be fully decoupled. 

In this work, we expand upon the ideas from this simple example and apply them to the DES Year 3 shear analysis. We introduce our formalism for accounting for blending in \sect{sec:shear_biases}.
In \sect{sec:thesims} we describe  realistic simulations of the DES Year 3 survey, validating them against the data. Then in \sect{sec:constant_shear} we describe and investigate the source of the traditional shear calibration biases estimated from constant shear simulations. In \sect{sec:zslice_shear}
we show that the biases described above, due to blended sources with different applied shears, are present in these DES Y3 simulations. We then present a method to 
combine mean shear measurements from the simulations with estimated redshift distributions in order to 
jointly infer corrections to both the shear calibration and the redshift distributions. We apply this method to the DES Year 3 simulations, producing a parameterized model of these effects that can be 
used to interpret the DES Year 3 shear catalogs, which we describe in \sect{sec:y3nz}. We summarize and discuss directions for future work in \sect{sec:conclusions}.

\section{Quantifying Shear Calibration Biases for Weak Lensing Shear Statistics}\label{sec:shear_biases}
In the following, we describe our formalism and methodology of using image simulations to calibrate gravitational lensing measurements. To this end it is useful to distinguish \emph{galaxies} from \emph{detections}. We take a galaxy to be emitting light \emph{of fixed redshift $z$}, with a particular surface brightness profile local to a position $\bm{\theta}$ on the sky. A detection on the other hand is, simply put, a thing 
identified by an algorithm designed to detect and deblend astronomical sources,  such as that employed by \textsc{SExtractor} \citep{sextractor}. Due to blending, measurements made on a detection may be affected by light from multiple
galaxies or stars and therefore multiple redshifts. 
We only have access to detections in an imaging survey, and we measure statistics averaged over ensembles of detections. We thus must determine how effects such as blending affect weak lensing shear statistics of ensembles of detections, rather than individual galaxies. 

In the following, we will work with the simplest such statistic, $\gobs(\bm{\theta})$, the mean measured shear over all detections within some pixel at some angular position $\bm{\theta}$. This is sufficient for our purposes since more cosmologically interesting 2-point statistics can be written as functions of such pixel-averaged mean shears, allowing the straightforward propagation of shear calibration biases, at least in the case that  spatial correlations of the biases can be neglected (see e.g. \citealt{kitching19} for an investigation of higher-order effects from relaxing this assumption).
We assume there is some true, redshift-dependent shear field $\gtrue(\bm{\theta},z)$. We can assume that in the weak lensing regime, where contributions of order $(\gtrue)^2$ or higher can be neglected, the mean measured shear is some linear function of the true shear field, 
\begin{equation}
    \gobs(\bm{\theta}) = \int_0^{\infty} \dx{z} \neff(z) \gtrue(\bm{\theta}, z) + c + \eta ,
    \label{eqn:ngamma}
\end{equation}
where $c$ is an additive bias, and $\eta$ is some measurement noise (due to e.g. the intrinsic shapes of galaxies, or pixel noise, that averages to zero over many such measurements). In the following we will often drop $\eta$ for compactness, such that $\gobs(\bm{\theta})$ is really the expectation value of the mean measured shear.
The response of the mean measured shear to the true shear field, which has traditionally been described as the multiplicative bias, $m$, is now described by the function $\neffz$, which we call the \emph{effective redshift distribution for lensing}, or \emph{effective redshift distribution} for short. In the next sections, we consider what  this function is in idealized conditions, and how to estimate it from image simulations under less idealized conditions. 

\subsection{Isolated galaxies}

It is useful to begin by considering the simplified case where each detection corresponds to a single galaxy. 
The image of each galaxy, and thus each detection, is subject to a gravitational shear field $\gtrue(\bm{\theta}, z)$, which depends only on its position $\bm{\theta}$ in the sky and on its redshift $z$. 
Consider an ensemble of 
detections, e.g.~a single photometric redshift bin in a gravitational lensing analysis. 
If each detection's shear is measured without bias, and each detection contributes to the mean with equal weight, then we have
\begin{equation}
    \gobs(\bm{\theta}) = \int_0^{\infty} \dx{z} n(z) \gtrue(\bm{\theta}, z). \label{eq:nz1}
\end{equation}
Here $n(z)$ is the derivative of the number of detections per unit area with respect to redshift
\begin{equation}
    n(z') \propto \left.\frac{\dx{N}}{\dx{z}}\right|_{z=z'}
\end{equation}
and is normalized to unity, such that no denominator is required in \eqn{eq:nz1}. Hence in this case, as expected the effective redshift distribution is simply what is usually called the \emph{redshift distribution} i.e. $\neffz = n(z)$.
One can estimate $n(z)$ as a finely sampled histogram of redshifts of individual galaxies associated with the detections,
\begin{equation}
    n(z) \Delta z \approx \frac{\text{\# of galaxies with $z<z_j<z+\Delta z$}}{N},
\end{equation}
where $N$ is the total number of galaxies, or
\begin{equation}
    n(z) \propto \sum_j \delta(z-z_j) \; . \label{eqn:nz_pz_sum}
\end{equation}

\subsection{Isolated galaxies with shear measurement biases}

Now suppose that for detections in the shear catalog (which we assume still have a one-to-one correspondence to galaxies) at redshift $z$, there is a mean multiplicative bias ($\bar{m}(z) \neq 0$), or equivalently non-unity mean response to shear ($\bar{R}(z) \neq 1$). This could be due to any imperfect correction of the  biases discussed in \sect{sec:intro} in the measurement of observed shear. The mean measured shear is now given by
\begin{equation}
    \gobs(\bm{\theta}) = \int_0^{\infty} \dx{z} \bar{R}(z)n(z) \gtrue(\bm{\theta}, z). \label{eqn:nz2}
\end{equation}
The effective redshift distribution is now given by
\begin{equation}
    \neffz = \bar{R}(z)n(z) \propto \bar{R}\frac{\dx{N}}{\dx{z}}
    \label{eqn:RN}
\end{equation}
and accounts not only for the fraction of galaxies in an ensemble at a given redshift, but also for how sensitive to shear our measurements of the shapes of those galaxies are (this will depend on the shape measurement method). For example, all galaxies at redshift $z'$ could be unresolved. Our catalog could contain some number of detections corresponding to them, whose shapes (and thus the shear they are subject to) we cannot measure. The mean observed shear of our catalog (or any other weak lensing statistic) would not respond to any true shear applied to light from $z'$. Thus, while these galaxies contribute to $dN/dz|_{z'}$, they should not contribute to $\neffz$.

The \mcal\ method provides estimates of the shear response for each detection, so it is straightforward to generalize \eqn{eqn:nz_pz_sum} to the case of non-unity response:
\begin{equation}
    \neffz = \frac{\sum_j R_j \delta(z-z_j)}{\sum_j R_j}\label{eqn:ngamma_sumRj},
\end{equation}
where $R_j$ is the \mcal\ shear response for detection $j$, and the use of $\sum_j R_j$ in the denominator ensures \neffz\ is normalized to unity. 

\subsection{Blended galaxies and the general case}\label{sec:ngamma}

The definition of $\neffz$ of \eqn{eqn:RN} seems natural and has been used not only in DES Year 1 \citep*{hoyle17}, but in some form by most other weak gravitational lensing analyses to date. It is accurate under the assumptions made -- that each detection corresponds to light from a single redshift, and that while response to shear may be imperfect, it can be described by its mean as a function of the redshift, or estimated based on some properties of the images of actual detections, each associated with a single redshift.

Let us now consider the less idealized, ubiquitous case of blending. Here, a detection is not always associated with a single galaxy. Rather, measurements on a single detection may be influenced by the light of multiple galaxies at different redshifts along the line of sight. The observed shape may respond to shear applied to any of these multiple galaxies.
It is not clear which galaxy's redshift to associate with the detection in order to estimate its contribution to $dN/dz$ and $\bar{R}(z)$. While methods exist for inferring distinct redshift components in blended systems \citep{Jones19,padmanabhan19}, in theory allowing the assignment of a detection to multiple redshifts, we still would not know the relative weight to assign to each redshift component in the ensemble $n(z)$. For a detection $j$, with components $k$ at redshift $z_k$, the correct weight $R_j^k$ is the linear response of the measured shape of the detection, $\gobs_j$, to a shear applied to component $k$ only, or equivalently at redshift $z_k$ only i.e.,
\begin{equation}
    R_j^k = \frac{\partial \gobs_j}{\partial g(z_k)}.
\end{equation}
The effective redshift distribution \neffz\ for the ensemble would be formed by summing over components $k$ for each detection $j$ i.e.
\begin{equation}
    \neffz = \frac{\sum_j \sum_k R_j^k \delta(z-z_k)}{\sum_j \sum_k R_j^k} \; . 
\end{equation}
Here $R_j^k$ is a generalization of the response used by \mcal, which is the response of the measured shape of a detection to a constant (i.e. redshift independent) shear. Unfortunately, unlike the \mcal\ response which can be computed numerically by applying an artificial shear to galaxy images, we cannot measure this generalization on real galaxy images, since it would require deblending the redshift components in order to shear them individually, which is impossible to do perfectly.

In the presence of blending, we can thus no longer assume a separable effective redshift distribution $\neffz=\bar{R}(z)n(z)$. Blending   however does not invalidate \eqn{eqn:ngamma}, which assumes only that in the  weak lensing regime, the observed mean shear \gobs\ can always be approximated as some linear function of the true shear field \gtruez. 
\Eqn{eqn:ngamma} in fact provides a definition of $\neffz$ via a functional derivative
\begin{equation}
    \neffz = \frac{\dx{\gobs}}{\dx{\gtrue}(z)} \;  .
    \label{eq:ngz1}
\end{equation}
One can also define \neffz\ in the limit $\Delta z \rightarrow 0$, 
\begin{equation}
    \neffz \Delta z = 
    \frac{\Delta \gbarobs} {\Delta{\gamma^{\text{true}}(z,z+\Delta z)}}.
    \label{eq:ngz2}
\end{equation}

We will later also use $N_{\gamma}(z_1,z_2)$, the integral of $n_{\gamma}(z)$ over some redshift interval $(z_1,z_2)$,
\begin{equation}
    N_{\gamma}(z_1,z_2) = \int_{z_1}^{z_2} \dx{z} \neff(z).
\end{equation}
This is the response of the mean measured shear of the ensemble $\bar{\bm{\gamma}}^{\text{obs}}$ to an applied shear $\Delta\gamma^{\text{true}}$ in redshift interval $(z_1,z_2)$.


\Eqn{eq:ngz1} and \eqn{eq:ngz2} constitute \emph{definitions} of the effective redshift distribution that should be used to describe a sample of detections in a weak gravitational lensing analysis e.g. for predicting shear correlation functions or tangential shear signals. 
Note that the normalization of $\neffz$, i.e. $\int_0^\infty dz \neffz$, is now meaningful. The normalization is the response of the mean measured shear to a true shear that is constant in redshift. This corresponds to the traditional $1{+}m$, where $m$ is the mean multiplicative bias for the ensemble. Note that numerical codes for making theoretical predictions of weak lensing statistics often internally normalize the provided redshift distribution. In this case one would need to apply the normalization of $\neffz$ to the predicted statistic as an additional correction. 

The use of $\neffz$ unifies two effects: biases in photometric redshifts and shear calibration, that have traditionally been treated separately. As we enter the era of deeper galaxy surveys, where blending becomes more and more important, we no longer have the luxury of treating these two systematic effects separately.

\subsection{Calibrating $\neff$ with image simulations}\label{sec:ngamma2}

The definition of \neffz\ in \eqn{eqn:RN} is superseded by that of \eqn{eq:ngz1}. In the presence of blending, the two are expected to disagree, not just in their normalization (i.e.~as an overall multiplicative bias), but also in their shape. 

Photometric and/or clustering-based redshift calibration, can at best (when combined with a \mcal\ response estimate for each detection) aim to measure $\bar{R}(z)n(z)$. Image simulations must be used to check whether the difference between $\bar{R}(z)n(z)$ and the true \neffz\ is small, as one might hope as long as blending is rare or mild. In case that difference is not negligible, image simulation methods can be used to infer corrections to $\bar{R}(z)n(z)$ that improve its agreement with the true \neffz; this is the approach taken in \sect{sec:zslice_shear} of this work. 

\Eqn{eq:ngz2} allows us to operationalize the updated definition of \neff. From suitable image simulations we can measure $N_{\gamma}(z_1,z_2)$, that is, the integral of \neffz\ over some interval $\alpha$ between $(z_1^{\alpha},z_2^\alpha)$, in which we vary the applied shear. To this end, we need to run a separate image simulation for each interval we would like to study, with a differential shear $\Delta\gamma^{\text{true}}_{\alpha}$ applied to galaxies within that interval. We will use more compact notation in the following, with $N_{\gamma}^\alpha$ denoting the integral of $\neffz$ over redshift interval $\alpha$, 
\begin{align}
N_{\gamma}^\alpha &\equiv N_{\gamma}(z_1^\alpha,z_2^\alpha) \\ 
&\equiv \int_{z_1^\alpha}^{z_2^\alpha} \dx{z} \neff(z) .
\end{align}
This is estimated from our simulations via
\begin{equation}
N_{\gamma}^\alpha =
\frac{\Delta \gbarobs}
{\Delta{\gamma^{\text{true}}_\alpha}},
\end{equation}
where $\Delta \gbarobs$ is the change in mean measured shear of the ensemble.

Thus far, we have only considered a single ensemble of detections, and its effective redshift distribution for lensing, $n_{\gamma}(z)$. In \sect{sec:zslice_shear}, we will estimate these quantities for multiple subsets of our simulated detections, for example true or photometric redshift bins. In this case, we assign these subsets a label $i$, and denote as $\gobs_i$, $\neffiz$ and $\Ngia$, the mean measured shear, effective redshift distribution and integrated effective redshift distribution  for the subset $i$ of our detections. 

In the absence of blending, and if there were no overlap between the redshift interval $\alpha$, and the redshift range of galaxies contributing to detections in subset $i$,  $\Delta \gbarobsi$ and thus $N_{\gamma,i}^\alpha(z)$ would vanish i.e. there would be no response of the mean shear for subset $i$ to shear in redshift interval $\alpha$. Blending, however, causes a non-zero response for any realistic selection of an ensemble $i$ of detections. This is because there will always be some amount of blending between galaxies in redshift interval $\alpha$, and the detections in $i$, and shear applied to the former will impact the measurement of the shapes of the latter.

\section{DES Y3 Image Simulations}\label{sec:thesims}

Having introduced and motivated our formalism for quantifying observational biases, we now turn to describing and validating our suite of DES Year 3-like image simulations. We note again here that DES Year 3 refers to the first three years of DES data processed together.
In our simulation design we follow closely the real DES Year 3 data, by simulating complete sets of single-epoch images required to form DES Year 3 \emph{tiles} in all four photometric bands $griz$, and then applying the same software for coadding, object detection and object measurement as is applied on the real data in \citet{y3-gold} and \citet*{y3-shapecatalog}. This consistent simulation of weak lensing data across multiple photometric bands is key a step forward in the Dark Energy Survey's joint shear and \photoz\ bias characterization.
We describe the main steps in our fiducial simulation pipeline below.

\subsection{Exposures, single-epoch images, and tiles}

DES wide-field images are processed in \emph{tiles}, square sky regions of side length $10,000$  pixel ($\approx0.72$ degree, see \citealt{morganson18}). There are 10,338 such tiles included in the Y3 dataset. For each tile region, all images
overlapping that region are included in a coadded image for each band. 
These input images to the coadd come from many \emph{exposures}, each of which is a collections of images, one for each CCD in the Dark Energy Camera (DECam, \citealt{flaugher15}) focal plane.  We refer to these individual CCD images as \emph{single-epoch images}.
We also organize our image simulation using this tiling system. 

We select at random 400 of the tiles entering the Y3 dataset, and for each tile, generate simulated versions of all the single-epoch images with any overlap of the tile region. For a given version of our simulation, this constitutes 62.8 million simulated objects, of which 15.4 million are detected in our shear pipeline, and 4.1 million pass shear catalog quality cuts (see \sect{sec:shear_est}). We simulate a total of 20 versions of this 400 tile set simulation, with versions differing only in their applied shear field (see \sect{sec:se_ims}), or whether objects are placed on a regular grid and whether detection is performed (see \sect{sec:simvariants}). We were limited in producing more simulation volume by time and computing resources, but find that this 400 tile set is sufficient volume such that statistical uncertainties in the simulation measurements are not the dominant uncertainty on our inferred bias corrections.


\subsection{Single-epoch Image generation}\label{sec:se_ims}
Our simulation pipeline starts by generating a simulated version of each single-epoch image using \textsc{GalSim}\footnote{\url{https://github.com/GalSim-developers/GalSim}} \citep{rowe14}, with the addition of various custom modules. Briefly, we create simulated images with noise, point-spread-functions and world-coordinate system (WCS) estimated from the DES Y3 data, and insert parametric models for stars and galaxies.
The images are simulated with the following properties:
\begin{itemize}
  \item{\textbf{Pixel geometry:} Each simulated single-epoch image is generated with the pixel geometry of the corresponding image in the real DES data. This is simply set by the DECam CCD properties, all of which have $4096\times2048$ pixels.}
  
  \item{\textbf{Noise:} Noise is assumed to be Gaussian and is drawn from the weight maps estimated for the corresponding image in the real DES data. For pixels that are masked in the corresponding image, the median of the weight map is used as the inverse noise variance. This noise field constitutes the only background onto which simulated objects are drawn --- we do not simulate e.g. non-zero sky background. This choice implicitly assumes that the background subtraction performed on the Y3 data is sufficiently accurate.}
  
  \item{\textbf{WCS:} Our input objects to the simulation, for which we generate a model in sky coordinates, are consistently drawn into all the single-epoch images they overlap, using the WCS (world-coordinate system) solution for the  the corresponding image in the real data.}
  
  \item{\textbf{PSF:} The drawn objects are convolved with a smoothed version of the PSF model estimated from the real data by Piff\footnote{\url{https://github.com/rmjarvis/Piff}} \citep{y3-piff}. See Appendix~\ref{app:piffmod} for the details and justification of this procedure. A significant simplification of our analysis here is that we do not attempt to measure the PSF from our simulations for use in shape measurement, rather we use the input PSF models. While we believe this is well justified by the stringent PSF validation performed in \citet{y3-piff}, jointly simulating the inference of PSF and shear would be a natural extension of the work presented here.}
  \item{\textbf{Masking:} We package with the simulated images the bad pixel masks taken from the corresponding real data image files. These indicate pixels to exclude or interpolate for downstream processing and measurement codes.}
  
  \item{\textbf{Input galaxies:} We randomly draw galaxy models from a catalog of "bulge+disk", two-component parametric fits to galaxies in the COSMOS field\footnote{\url{http://cosmos.astro.caltech.edu/page/hst}}. This catalog is described in \citet*{y3-deepfields}; we provide a brief description here. 
  The morphological parameters (half-light-radius, bulge-to-disc-ratio, ellipticity) of the parameteric galaxy models were fit to Hubble Space Telescope Advanced Camera for Surveys (HST-ACS) imaging \citep{scoville07,koekemoer07}, specifically we use the re-processed HST-ACS data of \citet{leauthaud07}. These model fits were then used to estimate fluxes for the DECam $griz$ filter bandpasses, using forced photometry at the same sky positions in deep stacks of DECam imaging (described in \citealt*{y3-deepfields}). The use of both HST imaging and DECam imaging gives us a catalog of parametric galaxy models with realistic and well-constrained morphology (from HST) and realistic fluxes and colors in the DES filters (from the deep DES imaging). When drawing a parametric model galaxy into the simulations, we apply a random rotation to each simulated object.
  
  We additionally match this catalog (using a 0.75 arcsecond matching radius) to the \citet{laigle16} redshift catalog, and  apply area masks for the problematic regions identified by \citet*{y3-deepfields}, after which 222116 unique input objects remain. We remove 60 ($0.03\%$) of these for which the parametric model fits failed. We additionally apply a selection $r_{50} > -0.25(\text{mag}_i - 22) - 1.35$, which we find effectively removes the stars (we separately simulate stars using a different catalog described below). 
  
  In our fiducial simulation we include only galaxies with $i$-band magnitude $<25.5$, which is two magnitudes fainter than our threshold for inclusion in the eventual shape catalog. While we do not perform an analysis of sensitivity to this choice, the results of e.g. \cite{hoekstra2017} suggest the absence of fainter objects than this in our simulations will probably bias our multiplicative bias estimates at the $\sim0.1\%$ level, well below our current uncertainties. 
  
  In each tile, the number of galaxies simulated is drawn from a Poisson distribution with mean 170000, which corresponds to a number density of 88 galaxies per square arcminute. The density of input objects was tuned such that the number of detected objects in the simulations matched the number of detected objects for the same set of tiles in the real DES Y3 data. The majority of these galaxies will be either undetected or removed via cuts, resulting in a number density of roughly 6 galaxies per square arcminute used for shear estimation.
  
  Galaxies are placed randomly on the sky and so are not clustered. Compared to the real Universe, we expect this to result in less blending of objects at similar redshifts. We discuss the implications of this approximation in Sections  ~\ref{sec:m_popreweight} and \ref{sec:conclusions}.
  }

  \item{\textbf{Input stars:} We use a catalog of stars simulated using the \textsc{Trilegal}\footnote{\url{http://stev.oapd.inaf.it/cgi-bin/trilegal}} code, with the best-fitting models from the \emph{MWFitting} method presented in \citet{pieres19}. This catalog contains both sky positions and fluxes for a simulated population of stars complete in the magnitude range 14-26 in the $g$-band. In our fiducial simulation we include only stars with $i$-band magnitude $<25.5$.}

  \item{\textbf{Input shear:} We run otherwise identical realizations of each simulation with constant input shears of either $+0.02$ or $-0.02$ in a single shear component (the other component being $0$).
  This allows us to follow the approach of \citet{pujol19}, who propose computing multiplicative shear biases via measuring the difference in recovered shear between two simulations which are identical apart from a small change in input shear. Using this procedure greatly reduces the noise (both shape noise and measurement noise) on the multiplicative bias estimate, since much of it cancels when taking the difference. Note this is closely related to the idea of the `ring-test' introduced by \citet{nakajima2007}. We additionally generate simulations where the applied shear depends on the redshift of the input galaxy. More specifically, we generate simulations where galaxies with redshift within some redshift interval $\alpha$ have a difference in applied shear of $\Delta \gtruea = 0.04$ with respect to the rest of the simulated galaxies. This allows us to measure the \Nga\ as described in \sect{sec:Ngamma_est}.}
\end{itemize}

\subsection{Image reduction, coaddition and detection}
The processing and measurements applied to the simulated images closely follows that performed on the real DES Y3 images (described in  \citealt{morganson18,y3-gold}; \citealt*{y3-shapecatalog}). We summarize here:
\begin{itemize}
\item{A $10000\times10000$ pixel weighted-mean coadd image is generated for each tile, in each band, using \textsc{SWarp} \citep{swarp}. The weight maps used are the same as those used to generate the noise on the simulated images.
The $r$, $i$ and $z$ coadd images are then themselves combined in a CHI-MEAN coadd image, again using \textsc{SWarp}.}
\item{The $riz$ coadd is used for object detection and segmentation, which is performed by \textsc{SExtractor} \citep{sextractor}. \textsc{SExtractor} outputs a catalog of detected objects, various measured quantities for these objects, as well as segmentation maps (which indicate which pixels in the images are assigned to which catalog objects).}
\item{Multi-epoch data structure (MEDS, \citealt{jarvis16}) files are generated for each band for each tile. For each detected object in the SExtractor catalog, the MEDS file contains a postage-stamp cutout from each of the single-epoch images in which that object appears. This data format makes convenient the fitting of models simultaneously to multiple observations of a given object.}
\end{itemize}

\subsection{Shear estimation}\label{sec:shear_est}
To generate a shear catalog for each tile, we run \mcal\ on the $r$, $i$ and $z$ MEDS files, which fits an elliptical Gaussian profile, convolved with the PSF model, to the observed light profile of each detection. The parameters of the profile are fit jointly to square regions (\emph{stamps}) extracted from each single-epoch image for all bands, apart from a free amplitude allowing an independent flux in each band. Stamps with a masked fraction of more than 0.1 are not used in the fits.
Estimates of the shear response, $R_{ij}$ are also generated by \mcal\, which is the response of the measured shear in component $i$ to an applied shear in component $j$. From the shear catalogs generated by \mcal\, we select a sample suitable for weak lensing measurements by applying the identical catalog cuts as those applied to the DES Y3 data in \citet*{y3-shapecatalog}. The most significant (in terms of number of objects removed) of these are cuts on the object signal-to-noise ratio, $\snr>10$, and the ratio of PSF-deconvolved galaxy size, $T$, to the PSF size, $T_{\mathrm{psf}}$, $T/T_{\mathrm{psf}}>0.5$. $T$ is an area measure (equal to the trace of the covariance) for the Gaussian profile. The signal-to-noise ratio cut is required to minimize biases associated with shear-dependence of the \sex\ selection, and the size cut reduces the impact of any PSF modeling errors. 
See \citet*{y3-shapecatalog} for more discussion of the motivation and details for the shear catalog cuts, as well as detailed descriptions of the quantities such as $T$. This sample can then be used to estimate the shear recovered from the simulation, which can then be compared to the true shear input to the simulation to estimate any biases in the shear recovery.

\subsection{Simulation variants}\label{sec:simvariants}

In order to better understand the source of shear calibration biases, we generate and analyse two sets of simulations additional to the \texttt{fiducial} simulation described thus far. In the \texttt{grid} simulations, 
objects are placed on a regular grid with spacing $\approx 35$ pixels ($\approx 9$ arcseconds). We would expect any biases related to blending to be absent in this variant. The second variant is the \texttt{grid-truedet} simulations, which again places objects on this regular grid, and in addition the \textsc{SExtractor} detection catalog used as input to the shape measurement is replaced by a catalog containing the true positions of input objects. We would expect this to remove any biases related to possible shear dependence of the \textsc{SExtractor} detection probability (e.g. if rounder objects were more likely to be detected). 

\subsection{Simulation Validation}\label{sec:simval}

The final part of our simulation pipeline is validating our simulation's realism by comparing it to the real data. 
We infer below that the shear biases present  are related to the blending of sources and the possible shear-dependence in how sources are selected, segmented and modelled. This blending must therefore be accurately characterized in the simulations. We ensure here that the number density of sources, the noise levels in the image, and the distribution of measured source properties like flux and size are well-matched between the fiducial simulation and real DES data. We expect the effects of blending to be sensitive to the properties of neighbouring objects (such as the number of them within a given distance, and their brightness) encountered by our target source galaxies, so we additionally study statistics sensitive to these in 
\sect{sec:m_popreweight}.





\begin{figure*}
\includegraphics[width=\columnwidth]{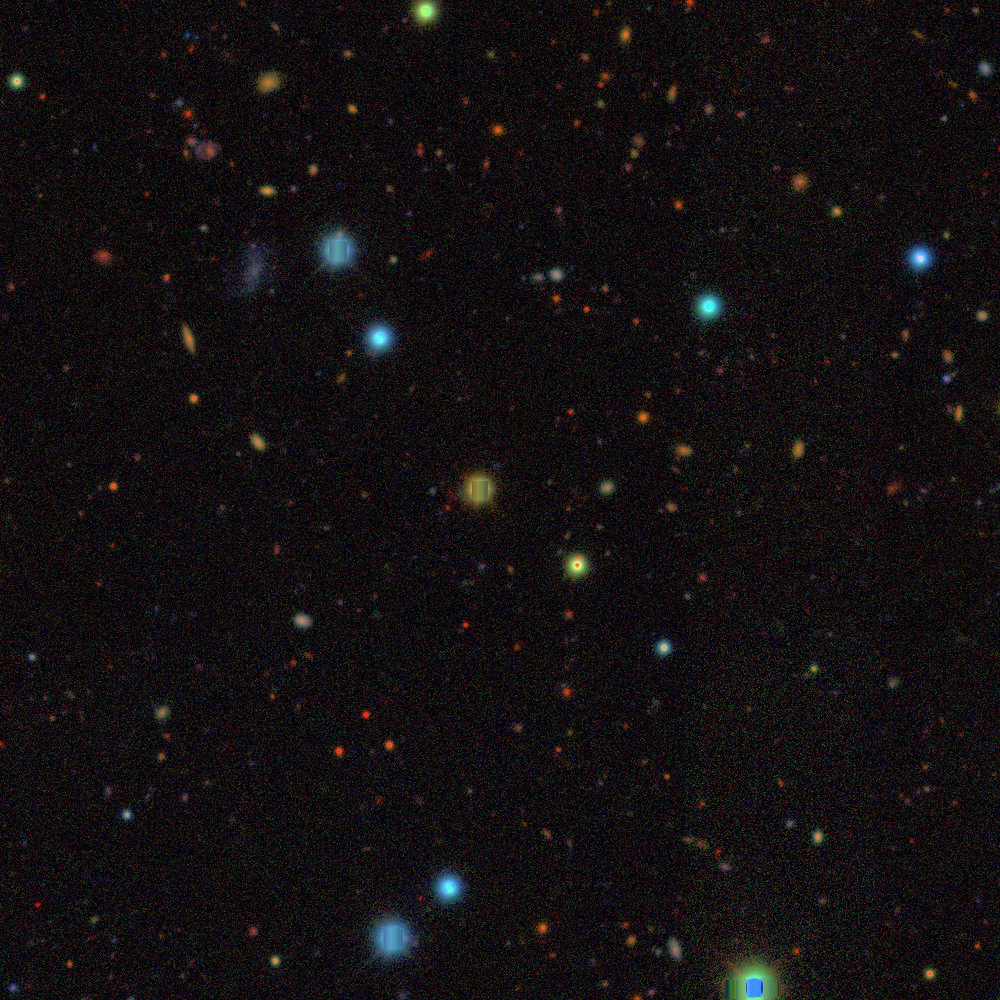}
\includegraphics[width=\columnwidth]{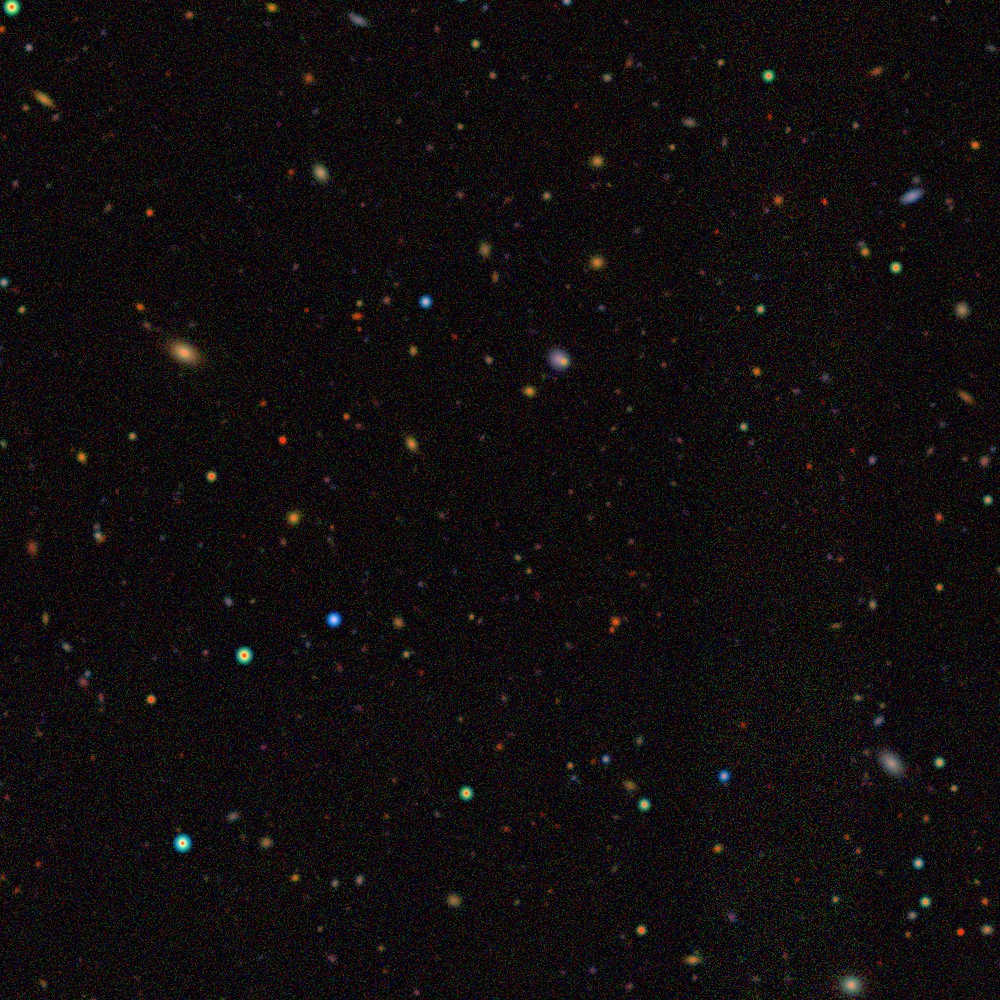}
\caption[]{$gri$ color image of $1000\times1000$ pixel region of tile DES0003-3832 for the real Y3 data (left panel) and the fiducial simulation (right panel).
.}
\label{fig:color_ims}
\end{figure*}

\subsubsection{Aesthetics} 
We generate $gri$ color images of our simulated coadds, using the \verb|desimage|\footnote{\url{https://github.com/esheldon/desimage}} package. These are useful for visual inspection of the simulations, but also for visual comparison to images made via the same process on the real data. \fig{fig:color_ims} shows color images for the same 1000x1000 pixel region of the coadd tile \verb|DES0003-3832| in the real data (left panel) and the fiducial simulation (right panel). A lack of bright stars in the simulation is apparent. These are masked out of the real data before analysis, so we do not believe that their absence from the simulations can impact our results. 


\subsubsection{\textsc{SExtractor} comparisons}
\label{sec:sizemag}
\sex\ forms a crucial part of our pipeline in detecting and segmenting objects in the coadd images, so it is important that the simulated images analysed using \sex\ resemble the real data.
Here, we perform comparisons of the measured properties of objects detected in the simulations. 

\begin{figure*}
\includegraphics[width=0.33\textwidth, trim=0.5cm 0.2cm 0.7cm 0.6cm, clip]{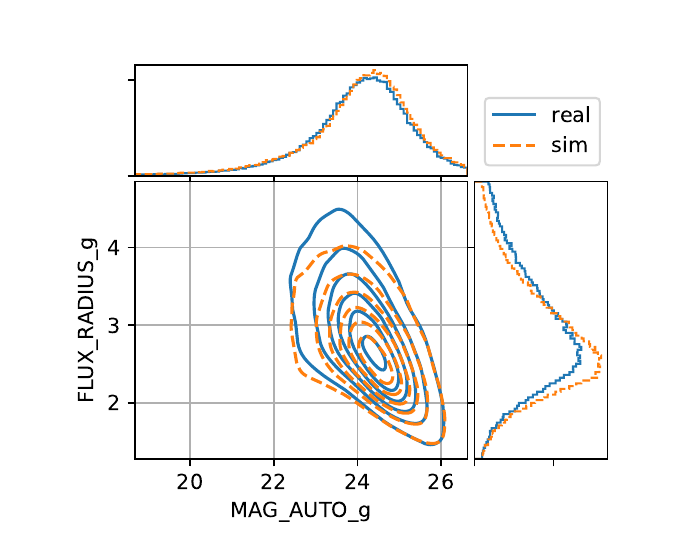}
\includegraphics[width=0.33\textwidth, trim=0.5cm 0.2cm 0.7cm 0.6cm, clip]{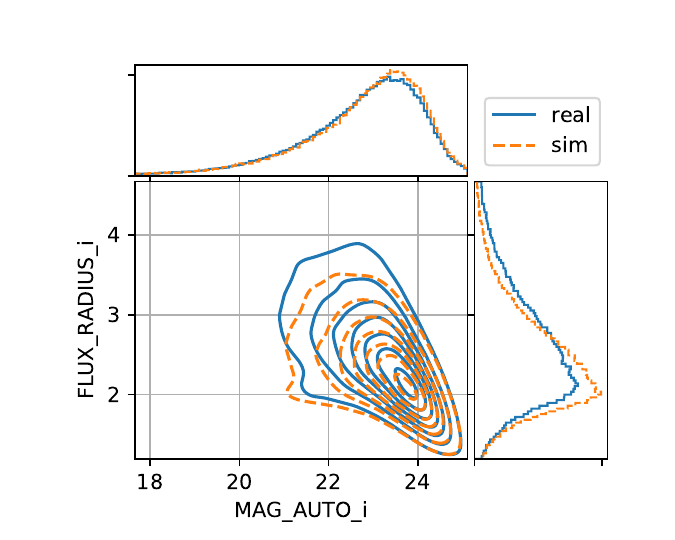}
\includegraphics[width=0.33\textwidth, trim=0.5cm 0.2cm 0.7cm 0.6cm, clip]{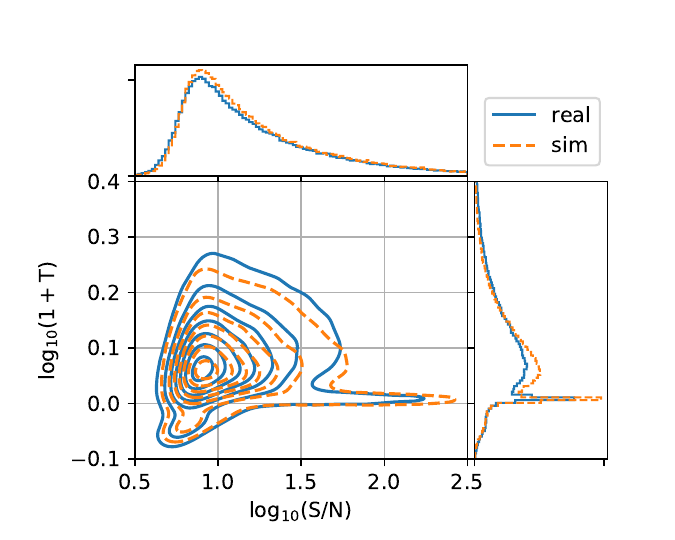}
\includegraphics[width=0.33\textwidth, trim=0.5cm 0.2cm 0.7cm 0.6cm, clip]{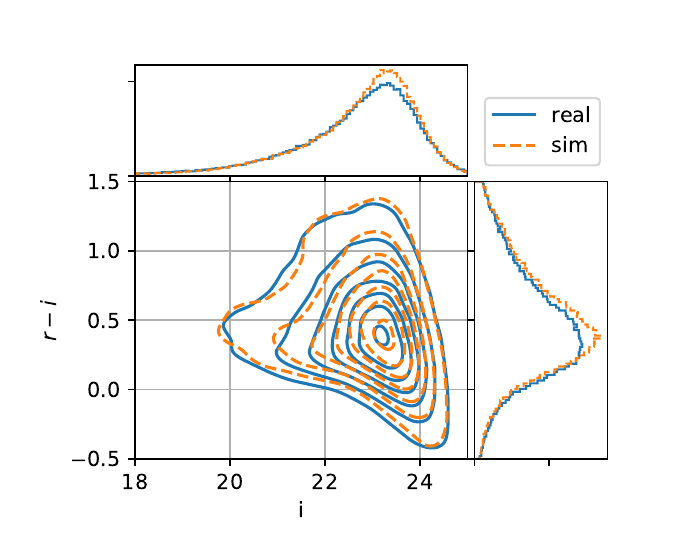}
\includegraphics[width=0.33\textwidth, trim=0.5cm 0.2cm 0.7cm 0.6cm, clip]{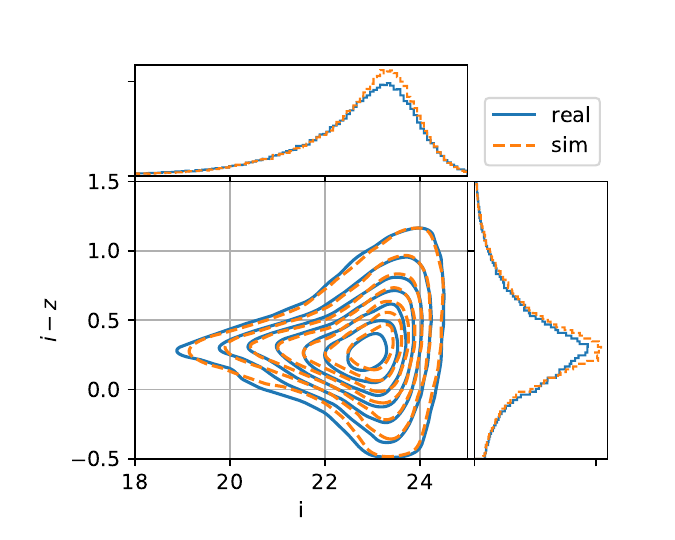}
\includegraphics[width=0.33\textwidth, trim=0.5cm 0.2cm 0.7cm 0.6cm, clip]{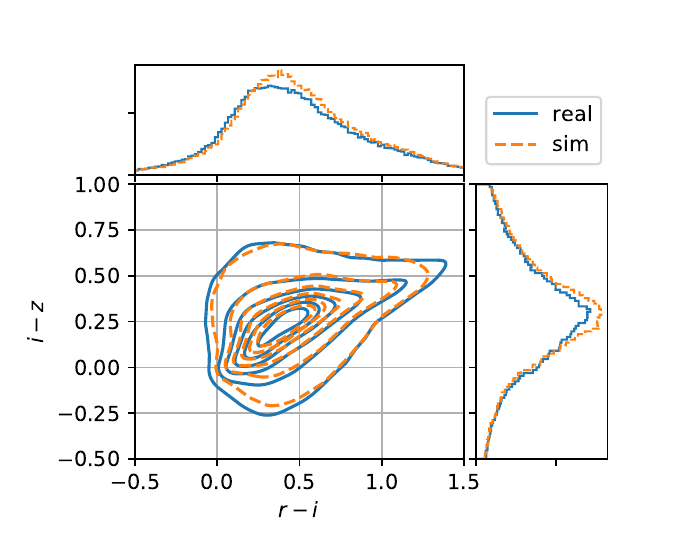}
\caption[]{Comparisons of joint distribution of measured quantities between simulations (orange dashed lines) and DES Y3 data (blue lines). The top-left and top-middle panels show joint
distributions of \sex\ measured quantities \texttt{MAG\_AUTO} and \texttt{FLUX\_RADIUS} $g$-band and $i$-band respectively. The stellar locus is at larger \texttt{FLUX\_RADIUS} in the $g$-band because of the larger PSF. The remaining panels show comparison of quantities estimated by the shape measurement code (\mcal). The top-right panel shows the joint distribution of  signal-to-noise and size (in fact, $\log_{10}(\snr)$ and $\log_{10}(1+T)$). The bottom-left panel shows the joint-distribution of $i$-band magnitude and $r-i$ color. The bottom-middle panel shows the joint distribution of $i$-band magnitude and $i-z$ color. The bottom right panel shows the joint distribution of $r-i$ and $i-z$ colors.}
\label{fig:mcal_comp}
\end{figure*}

The top-left and top-middle panels of \fig{fig:mcal_comp} shows the joint distribution of magnitude and size estimated by \sex, for the $g$ and $i$-bands respectively. Specifically, we use 
\verb|MAG_AUTO| (an elliptical aperture magnitude), and \texttt{FLUX\_RADIUS} (with $\verb|PHOT_FLUXFRAC|=0.5$) an estimate of the PSF-convolved half-light radius of the object\footnote{see \sex\ documentation at \url{http://mensa.ast.uct.ac.za/~holwerda/SE/Manual.html} or \url{https://www.astromatic.net/pubsvn/software/sextractor/trunk/doc/sextractor.pdf} for more details on these quantities}. Some clear features are apparent in both simulations and data, such as the shift of the distribution to larger size in the $g$-band compared to the $i$-band, which is due primarily to the larger PSF. We note that the distributions plotted share the same normalization, so any differences in absolute number density would be apparent.

\subsubsection{\mcal\ quantities}

For shear estimation and photometry, we use elliptical Gaussian models fit to single-epoch images (as opposed to the coadds used for the \sex\ quantities discussed above). The distributions of size and signal-to-noise of a weak lensing galaxy sample have often been identified as key to the expected level of bias in the shear recovery (e.g. \citealt{refregier12,kacprzak12}). The top-right panel of \fig{fig:mcal_comp} compares the joint distribution of signal-to-noise and size between simulations and data. The distributions are smoother in the log of these quantities, so we actually compare distributions of $\log_{10}(\snr)$ and (since $T$ can take slightly negative values) $\log_{10}(1+T)$.

As well as size and signal-to-noise, it is important to verify that other quantities used to select sub-samples of the source galaxies are well matched between simulations and data. A key example of this is selection of objects into bins based on their photometric redshift (\photoz\ henceforth). Photo-$z$ algorithms usually use some discrete set of broadband fluxes, with ratios of those fluxes (or differences in magnitudes), known as \emph{colors} considered particularly informative for redshift estimation. The bottom three panels of \fig{fig:mcal_comp} demonstrate that the simulations reproduce well the joint distributions of (from left to right): $i$-band magnitude and $r-i$ color, $i$-band magnitude and $i-z$ color, and $r-i$ color and $i-z$ color.



\subsection{Photo-z inference}\label{sec:sompz}

The photometric redshift distributions for the simulation sample are inferred using the \emph{SOMPZ} method outlined in \cite*{y3-sompz} (see also \citealt*{buchs19}). Following the methodology applied to DES Y3 data there, detected objects in the image simulations are assigned to two self organizing maps (SOMs), corresponding to ``wide'' (32 x 32 cells, using $riz$ flux information) and ``deep'' (64 $\times$ 64 cells, using $ugrizJHK_s$ colour information from the DES Deep Fields measurements presented in \citealt*{y3-deepfields}).  In the Y3 data anaysis, galaxies with  spectroscopic redshifts or  flux measurements in a large number of photometric bands (``deep" galaxies) are assigned to the higher resolution SOM. When those same galaxies are assigned to the lower resolution SOM based on their wide field flux information, they can be used to infer information about other galaxies with only the limited wide-field photometry available in order to calibrate the color-redshift relation. 
 
We use the SOMs constructed from the Y3 data catalogs and the same software to assign the simulated galaxies \citep*{y3-sompz}. We assume that all simulation input galaxies have precise redshifts (equal to the \photoz\ point estimates by \citealt{laigle16}) that, together with their measured deep $ugrizJHK_s$ fluxes,  describe the deep color-redshift relation. 
Matching simulation detections to input galaxies allows us to generate a transfer function that connects redshift, deep SOM cell, and wide SOM cell. The wide cells and their contained galaxies are then grouped together by their mean redshift in a variety of ways (see below) to form tomographic bins. 

The procedure for matching injection galaxies to measured detections in these simulations begins with a nearest neighbors search to identify the three closest objects in the truth catalog to a given detection. For a detection to be matched to a true object, it must have a close match within two pixels. If it does not have one, it is ignored in the rest of the \photoz\ inference (roughly $0.5\%$ of detections). Detections with an exclusive close match (i.e. one that is not a close match to any other detection) make up approximately $30\%$ of the simulated sample. To discriminate between detections with multiple close truth matches, we loop over injections from brightest to faintest in $i$-band magnitude and assign the brightest close truth match that has not yet been assigned to another detection. Roughly $70\%$ of detections fall into this category. If all close matches have already been assigned to other detections via this loop, no truth match is assigned, but this happens rarely enough to be negligible. For example, this only occurs for 4 cases in the entire fiducial $(g_1,g_2)=(-0.02,0.00)$  simulation.  

\begin{figure*}
    \centering
    \includegraphics[width=1.0\textwidth]{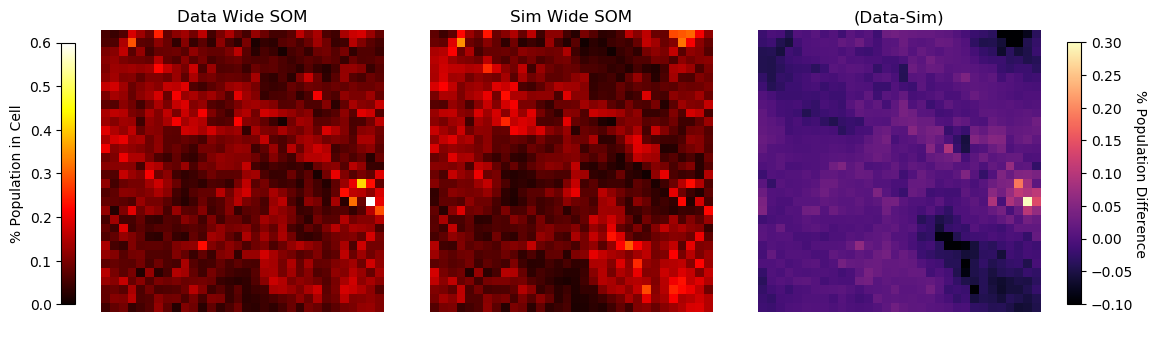}
    \caption{The wide SOM population distribution in the data (left) compared to the simulations (center), with a residual (right) that shows the majority of cells agree to within 0.05\% in total population.}
    \label{fig:som_occup}
\end{figure*}

   
The wide SOM occupancy for both the simulations and the data can be seen in Figure \ref{fig:som_occup}, showing very good general agreement. This is a consequence of the close alignment of color and magnitude distributions between the data and the simulations discussed in Section \ref{sec:sizemag}. The largest point of discrepancy is due to cells composed of very large, very blue galaxies at low redshifts, which likely do not significantly contribute to the redshift-dependent effects of blending that are key to this analysis.


A principal aim of this work is to quantify the shear calibration separately for each of the photometric redshift bins used in the DES Y3 cosmology analyses. This requires that we perform an equivalent photometric redshift binning on the image simulations. 
One can motivate several different choices of \photoz\  binning algorithms in order to estimate the shear calibration biases that may differ due to the small differences in color-redshift relation between the sims and the data. Presented here is a description  of the four binning algorithms we use and their resulting summary statistics like the mean redshift per bin.

\begin{itemize}
    \item \textbf{Fiducial}: We use the same mapping between wide SOM cell and \photoz\ bin as in the Y3 data. We make this our fiducial choice, since we think the philosophy of treating the real data and the simulations as similarly as possible is a sensible one. However,  small differences in wide SOM occupancy between simulations and real data are apparent with this procedure. Whereas in the data, each \photoz\ bin has an equal number of objects (25\% of the total, by construction), there are deviations from 25\% occupancy in the simulations, with the four \photoz\ bins receiving 19\%, 24\%, 25\% and 30\% of the objects respectively.
    \item \textbf{Equal Count (\texttt{equal})}: Instead of taking the mapping from the data, the wide SOM cells are ordered by mean redshift and then grouped such that an approximately equal number of galaxies end up in each \photoz\ bin, much like how the mapping is chosen in the data.
    \item \textbf{Mean redshift  matching (\texttt{z-match})}: 
    This is a binning that chooses the wide SOM cells such that they closely match the mean redshift found in the data \photoz\ bins, with approximately equal counts in each bin. By necessity, in this binning scheme a galaxy may end up assigned to multiple bins, or not assigned to any bin.
    \item \textbf{SOM Occupancy Matching (\texttt{w-match})}: This preserves the mapping used in the fiducial case, but also reweights galaxies to reproduce the relative wide SOM cell occupancy found in the Y3 data, using the ratio between the left and center panels of \fig{fig:som_occup}.
\end{itemize}

\begin{figure*}
    \centering
    \includegraphics[width=0.8\textwidth]{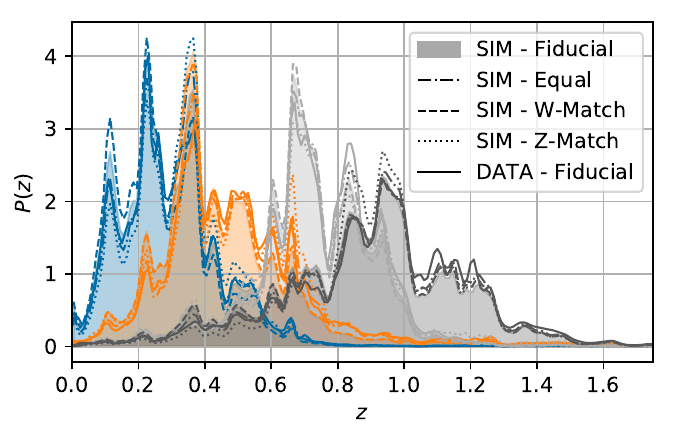}
    \caption{The photometric redshift distributions for the four \photoz\ binning schemes applied in the simulations (described in \sect{sec:sompz}, as well as the estimated Y3 data redshift distributions (labelled ``DATA - Fiducial''). For plotting purposes all curves have been smoothed using a Gaussian kernel of width 0.01 in $z$.  
    }
    \label{fig:Nzs}
\end{figure*}

The mean redshift for each of these \photoz\ binning choices on the simulations are reported in \tab{tab:meanz}, and the full distributions can be seen in \fig{fig:Nzs}. We see broad similarity between the binning recipes, but revisit the alternative options in \sect{sec:m_pzbinning} where we test the sensitivity of our calibration corrections to the choice of binning recipe.

\begin{table}
\begin{tabular}[width=0.9*\textwidth]{ c|c|c|c|c|c }
 \hline
 Sample & Binning Style & $\bar{z}_0$ & $\bar{z}_1$ & $\bar{z}_2$ & $\bar{z}_3$\\ 
 \hline
 Data & Fiducial & 0.334 & 0.517 & 0.749 & 0.936\\ 
 Sim & Fiducial & 0.311 & 0.460 & 0.723 & 0.894\\ 
 Sim & Equal & 0.322 & 0.510 & 0.745 & 0.920\\ 
 Sim & Z-Match & 0.325 & 0.511 & 0.753 & 0.930\\  
 Sim & W-Match & 0.294 & 0.463 & 0.714 & 0.895\\   
 \hline
\end{tabular}
\caption{\label{tab:meanz} Table describing the mean redshift per \photoz\ bin for each binning algorithm as compared to that found in the data.}
\end{table}


\section{Results I: Shear calibration biases from constant shear simulations}\label{sec:constant_shear}

In this section, we begin by examining the shear calibration biases apparent in constant shear simulations, and we report average shear calibration bias estimates for the full DES Y3-like sample, as well as for individual \photoz\  bins. As described in \sect{sec:shear_biases}, these bias estimates are not sufficient in general to correct theoretical predictions for weak lensing shear statistics, hence in \sect{sec:zslice_shear}, we present estimates of biases in \neffz, using simulations where the input shear varies with redshift. 

For these calculations, we use four 400 tile constant shear simulations with input shear
\begin{equation}
(g_1,g_2) \in \left\{(0.02,0), (-0.02,0), (0,0.02), (0,-0.02)\right\}
\end{equation}
(see \sect{sec:se_ims}).
Each simulation is identical (i.e. random elements of the simulation use the same random seeds) apart from the applied shear,  reducing the noise on the differences between the mean shear measured between pairs of the simulations (see \citealt{pujol19}). See \tab{tab:simulations} for more details. 

For each shear component $\mu$, the multiplicative and additive biases, $m_{\mu}$ and $c_{\mu}$ are calculated by fitting the model
\begin{equation}
    \gobs_x = (1+m_x)\gtrue_x + c_x
\end{equation}
to the pair of simulations with $\gtrue_x = \pm 0.02$, where $\gobs_x$ is the measured mean shear for component $x$. The uncertainty on $\gobs_x$, inferred from jackknifing over the simulated tiles, is included in the fit.

\begin{figure}
\includegraphics[width=0.9\columnwidth]{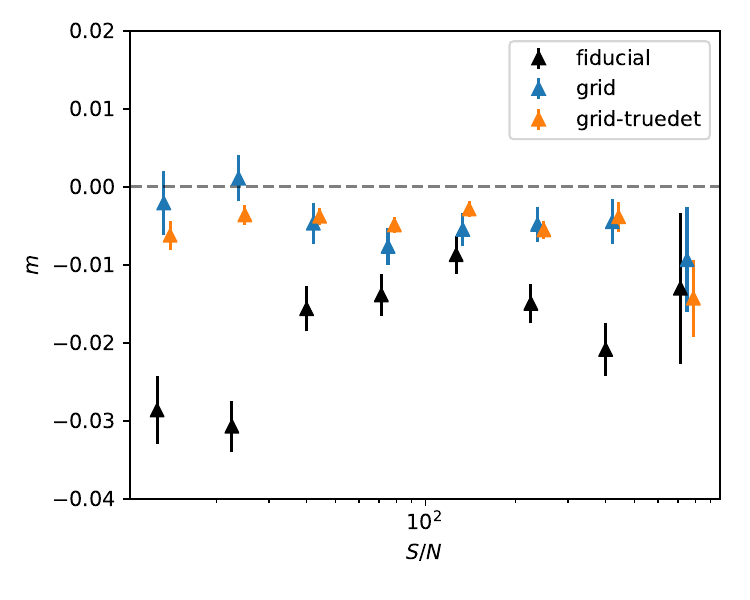}
\caption[]{
  Multiplicative bias as a function of \snr\ for the \fiducial, \grid, and \gridtruedet\ simulations. The \fiducial\ simulation places 
  the objects in the images at random positions and detects them with \sex. The \grid\ simulation places them on a grid with $\approx9$ arcsecond 
  spacing and still employs \sex. The \gridtruedet\ simulation uses a grid for the objects, but uses their true positions instead of detecting 
  them with \sex.}
\label{fig:m_v_snr}
\end{figure}

\tab{tab:mean_m} contains mean multiplicative biases for our \texttt{fiducial}, \texttt{grid}, and \texttt{grid-truedet}  simulations. In the \texttt{grid} simulation, blending is removed by placing down objects on a grid. In the \gridtruedet\ simulation, objects are again placed on a grid, and in addition we use their true positions for the detection catalog, rather than the detection catalog  estimated using \sex. We expect this to additionally remove any selection biases due to shear-dependence of the \sex\  selection. 

While the fiducial simulations exhibit a mean multiplicative bias of $\approx-2\%$, the multiplicative bias for both the \grid\ and the \gridtruedet\ simulations are greatly reduced, with a remaining bias of around $-0.4\%$. We can draw a few conclusions from this. Firstly, the fact that we see consistent biases for the \grid\ and \gridtruedet\ cases implies that we do not have significant \sex\ biases for isolated objects. This is not unexpected, since we apply a $\snr>10$\ cut to our catalogs that likely dominates over the threshold for detection used by \sex, and \mcal\ is able to accurately correct for selection biases due to this cut.

Beyond that, we can attribute most of the multiplicative bias we see in the fiducial simulation to the presence of blending. As explored in \citet{sheldon2020}, the presence of blending can likely generate biases through various (related) mechanisms. Firstly there is `detection' bias due to shear-dependence of the detection algorithm, including the decision of how many different objects to assign to a blend. Secondly one may also expect an additional bias due to the presence of a neighbor in the shape measurement process, either due to its contaminating flux, or due to some unaccounted-for shear dependence of the neighbor masking algorithm. With these simulations, we cannot fully decouple these effects. A simulation with randomly-placed objects as in the fiducial simulation, but using true detection to remove detection biases, would shed light on the issue, but we did not have the resources to run this variation.

\fig{fig:m_v_snr} shows the multiplicative bias (averaged over shear component) as a function of signal-to-noise ratio. For the \grid\ and \gridtruedet\ cases there is no clear trend with $\snr$, while for the fiducial case, there do appear to be variations with $\snr$, although only at the percent level, and we do not attempt to explain this behaviour. We also do not have a conclusive explanation for the $\approx -0.4\%$ biases remaining in the \grid\ and \gridtruedet\ simulations. Appendix~\ref{app:simval} explores various potential sources of small (sub-percent) multiplicative biases using idealized simulations, and we do see the potential for biases due to masking at the $\sim0.1\%$, so this may be contributing some of the remaining bias. Given that we directly apply the masks from the DES Y3 data to our simulations, we are confident that the biases in the simulations due to masking will be at a similar level to that present in the real data. 

\begin{figure}
\includegraphics[width=0.9\columnwidth]{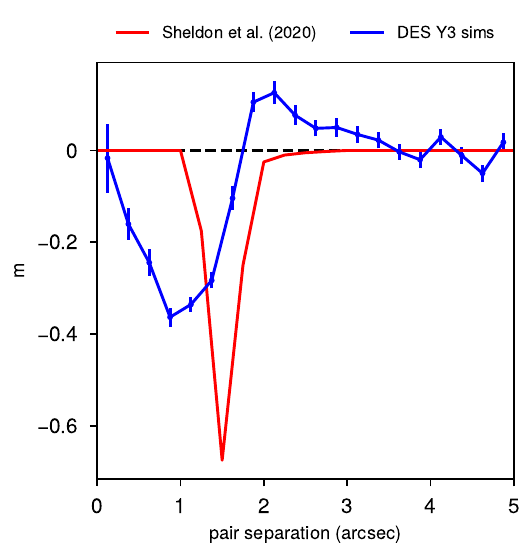}
\caption[]{
  Multiplicative bias for detections corresponding to pairs of truth objects separated by a given angle. To make this plot, we find pairs of true objects separated by some distance in the truth catalog in our fiducial simulations and not near any other objects. We then use detected objects near this location to measure the multiplicative bias as a function of separation. At small scales, we find signals in our simulations that appear to be from detection bias. For comparison, we show the detection bias result from \citet{sheldon2020} as the red line. At intermediate scales, we see a positive bias, which is probably from pairs where one source is much fainter than the other.
  The bias converges to be consistent with zero at large separations.}
\label{fig:pairsmtheta}
\end{figure}

In order to provide some more detail on the impact of blending on the multiplicative bias, we attempt to compute the bias as a function of object separation as follows. Using the galaxies in the input truth catalogs for our simulations, we find pairs of input galaxies that are separated by some distance and are no closer to any other galaxy in the simulation than the other galaxy in its pair.
We then define a small region in the simulated image encompassing each pair, such that detections in this region include light from just these two galaxies.  Specifically, we used a circle of radius 0.75 times the pair separation, centered on the midpoint between the two galaxies. Detections in these regions were used to compute the multiplicative bias as a function of separation. The results of this exercise are shown in Figure~\ref{fig:pairsmtheta}. For comparison, we show the predicted effects of detection bias from \citet{sheldon2020} for a simplified simulation setup involving only pairs of galaxies with equal size and flux. Note also that \citet{sheldon2020} used a  different procedure to remove the light of neighboring objects. At small separations we see the qualitative effects of detection bias, that is a large negative bias. At large separations, the bias is consistent with zero within the statistical noise. At intermediate scales, we also see a positive bias that was not apparent in the \citet{sheldon2020} simulation results. This can occur when a neighbor is not detected, so may only be significant when pairs with widely disparate fluxes are included.

To summarize, we see multiplicative biases in our simulations from at least three sources. First, as shown in Appendix~\ref{app:simval}, the effects of masking and their corrections used in the DES Y3 analysis cause a small, few tenths of a percent bias. Second, for very close pairs of truth objects, we see a strong negative multiplicative bias in \fig{fig:pairsmtheta} that is qualitatively consistent with the detection biases studied in \citet{sheldon2020}. At intermediate separations we see some positive multiplicative bias which is presumably blending-related, involving pairs with widely different fluxes. 

We note here that the mean multiplicative bias of $\approx-0.02$ we estimate here is somewhat different from the multiplicative bias prior of $0.012\pm 0.012$ inferred by \citet*{zuntzsheldon18} for the DES Year 1 shear catalog, which used very similar shape measurement methodology. We believe the significant improvements in simulation realism presented here explain this discrepancy. We note also that \citet{sheldon2020}, which used an independent set of image simulations to those presented here, that we believe also contained significant improvements with respect to those used by \citet*{zuntzsheldon18}, reported multiplicative biases much close to those presented here for DES-like simulations.


In \sect{sec:m_popreweight}, we use a re-weighting procedure to estimate systematic uncertainty (due to potential simulation inaccuracy) in the impact of blending, which is then propagated to our final priors on the shear calibration.

\begin{table*}
\begin{tabular}[width=0.9*\textwidth]{ c|c|c|c|c|c }
 \hline
 variant & sheared redshift interval & $(g_1,g_2)$ in redshift interval & $(g_1,g_2)$ outside redshift interval & object placement & SExtractor detection \\
 \hline
 \gridtruedet   & [0.0, 3.0] & $(+0.02, 0.00)$ & -- & grid   & no\\ 
 \gridtruedet   & [0.0, 3.0] & $(-0.02, 0.00)$ & -- & grid   & no\\ 
 \gridtruedet   & [0.0, 3.0] & $(0.00, +0.02)$ & -- & grid   & no\\ 
 \gridtruedet   & [0.0, 3.0] & $(0.00, -0.02)$ & -- & grid   & no\\   
 \hline
 \grid           & [0.0, 3.0] & $(+0.02, 0.00)$ & -- & grid   & yes\\ 
 \grid           & [0.0, 3.0] & $(-0.02, 0.00)$ & -- & grid   & yes\\ 
 \grid           & [0.0, 3.0] & $(0.00, +0.02)$ & -- & grid   & yes\\ 
 \grid           & [0.0, 3.0] & $(0.00, -0.02)$ & -- & grid   & yes\\  
 \hline 
 \fiducial       & [0.0, 3.0] & $(+0.02, 0.00)$ & -- & random & yes\\ 
 \fiducial       & [0.0, 3.0] & $(-0.02, 0.00)$ & -- & random & yes\\ 
 \fiducial       & [0.0, 3.0] & $(0.00, +0.02)$ & -- & random & yes\\ 
 \fiducial       & [0.0, 3.0] & $(0.00, -0.02)$ & -- & random & yes\\ 
 \hline
 \fiducial       & [0.0, 0.4] & $(+0.02, 0.00)$ & $(-0.02, 0.00)$ & random & yes\\ 
 \fiducial       & [0.4, 0.7] & $(+0.02, 0.00)$ & $(-0.02, 0.00)$ & random & yes\\ 
 \fiducial       & [0.7, 1.0] & $(+0.02, 0.00)$ & $(-0.02, 0.00)$ & random & yes\\ 
 \fiducial       & [1.0, 3.0] & $(+0.02, 0.00)$ & $(-0.02, 0.00)$ & random & yes\\ 
 \fiducial       & [0.0, 0.4] & $(0.00, +0.02)$ & $(0.00, -0.02)$ & random & yes\\ 
 \fiducial       & [0.4, 0.7] & $(0.00, +0.02)$ & $(0.00, -0.02)$ & random & yes\\ 
 \fiducial       & [0.7, 1.0] & $(0.00, +0.02)$ & $(0.00, -0.02)$ & random & yes\\ 
 \fiducial       & [1.0, 3.0] & $(0.00, +0.02)$ & $(0.00, -0.02)$ & random & yes\\ 
 \hline 
\end{tabular}
\caption{\label{tab:simulations} Image simulation properties. Each line lists a specific image simulation produced for this work. The \fiducial simulation places objects at random into the image and uses \sex\ for object detection. We also ran variations for some of the simulations where objects were placed on a grid (\texttt{grid}) or we used their true locations for shear measurements (\texttt{grid-truedet}).}
\end{table*}

\begin{table*}
\begin{tabular}[width=0.9*\columnwidth]{ c|c|c|c|c|c }
 \hline
 variant & $m_1 \times 100$ & $m_2 \times 100$ & $m \times 100$ & $c_1\times10^4$ & $c_2\times10^4$ \\ 
 \hline
 \texttt{grid-truedet} & $-0.48 \pm 0.07$ & $-0.40 \pm 0.07$ & $ -0.44 \pm 0.05$ & $-2.06 \pm 1.48$ & $-2.77 \pm 1.46$ \\ 
 \texttt{grid} & $-0.33 \pm 0.15$ & $-0.35 \pm 0.15$ & $-0.34 \pm 0.11$ & $-2.14 \pm 1.34$ & $-3.59 \pm  1.40$ \\ 
 \texttt{fiducial} & $ -2.23 \pm 0.16$ & $-1.93 \pm 0.17$ & $-2.08 \pm 0.12$ & $-0.86 \pm 1.35$ & $-1.34 \pm 1.45$\\ 
 \texttt{fiducial} bin 0 & $-1.55 \pm 0.44$ & $-0.96 \pm 0.45$ & $-1.25 \pm 0.31$ & $1.57 \pm  3.05$ & $-4.38 \pm 2.75$ \\
 \texttt{fiducial} bin 1 & $-1.77 \pm 0.56$ & $-1.87 \pm 0.54$ & $-1.82 \pm 0.39$ & $-1.86 \pm  2.66$ & $-0.69 \pm 2.79$\\
 \texttt{fiducial} bin 2 & $-2.51 \pm  0.64$ & $-2.03 \pm 0.67$ & $-2.27 \pm 0.44$ & $-3.05 \pm 2.39$ & $1.11 \pm  2.43$\\
 \texttt{fiducial} bin 3 & $-0.038 \pm 0.79$ & $-0.034 \pm 0.88$ & $-3.60 \pm 0.59$ & $0.31 \pm 2.78$ & $-1.24 \pm  2.89$\\
 \hline
\end{tabular}
\caption{\label{tab:mean_m} Average multiplicative ($m$) and additive ($c$) biases for the \texttt{fiducial}, \texttt{grid} and \texttt{grid-truedet} simulations. While we expect no bias in the shear measurements for simulations of objects on a grid with no detection employed, we do find a small bias due to masking corrections. See Appendix~\ref{app:simval} for details. The \texttt{fiducial} simulations show non-trivial multiplicative biases due to a combination of blending, object detection effects, masking, 
and potentially other unknown causes.}
\end{table*}

\section{Results II: : Estimates of \lowercase{$n_{\gamma}(z)$} biases from redshift-dependent shear simulations}\label{sec:zslice_shear}

We explained in \sect{sec:shear_biases} that for theory predictions of shear statistics involving ensembles of detections over a range of redshifts, we require an estimate of $n_{\gamma}(z)$, the effective redshift distribution for lensing. In this section we describe the extra simulations and methodology used to infer biases in the DES Y3 methodology for estimating $n_{\gamma}(z)$. To summarize, we take the following approach (with much more detail given in the following section). For each \photoz\ bin $i$:
\begin{itemize}
    \item We measure $N_{\gamma,i}^{\alpha}$ for 4 redshift intervals $\alpha$ (with ranges $(z_1^\alpha,z_2^\alpha)$), from simulations with a change in applied shear within that interval. We use redshift intervals $(z_1^\alpha,z_2^\alpha) \in \{(0.0,0.4),(0.4,0.7),(0.7,1.0),(1.0,3.0)\}$. Note these redshift intervals $\alpha$ have no specific correspondence to the four \photoz\ bins $i$ we use.
    \item We compare this measurement to the prediction based on integrating the \mcal\ response-weighted redshift distribution, $\neffzmcal$ (which is how $\neffz$ is estimated in the real data).
    \item Due to computing limitations, we can only measure $N_{\gamma,i}^
    {\alpha}$ in the four coarse aforementioned redshift intervals. However, we need a finely sampled $\neffz$ to make theory predictions. Therefore we use a model that parameterizes deviations from \neffzmcal\, of the form $n_{\gamma}^{\text{model}}(z) = f(z)\neffzmcal + g(z)$, where $f(z)$ and $g(z)$ are smooth functions of $z$. We fit this model to our $N_{\gamma,i}^
    {\alpha}$ measurement.
\end{itemize}

We start in \sect{sec:Ngamma_est} by describing the simulation inputs and procedure for estimating $\Nga$ for a given ensemble of detections. In \sect{sec:Ngamma_nontomo} we present the $N_{\gamma,i}$ estimates for the case without redshift binning, and compare these to the prediction from the \mcal\ response-weighted redshift distribution, $\neffzmcal$.
In \sect{sec:Ngamma_slice} we present measurements of $N_{\gamma,i}^\alpha$ for ensembles of galaxies restricted to true redshift intervals, a case which most clearly demonstrates the response of galaxies assinged one redshift to a shear applied at another. 
In \sect{sec:Ngamma_tomo} we present measurements of $N_{\gamma,i}^{\alpha}$, for each \photoz\  bin $i$, again comparing these to the $\neffzmcal$. In  \sect{sec:ngammamodel} we describe our modeling approach using fitting functions to infer \neffz\ corrections from these measurements, and finally in \sect{sec:model_constraints} present the resulting \neffz\ bias model constraints.

\subsection{Estimating \Nga}\label{sec:Ngamma_est}

As discussed in \sect{sec:shear_biases}, we can  estimate \Nga, which is \neffz\ integrated over the interval $z_1^\alpha<z<z_2^\alpha$, by generating a simulation which we can label $\alpha$, that has constant true shear $g^{\text{const}}$ apart from in redshift interval $\alpha$ which has an applied true shear  $\gconst + \Delta \gtruea$. Our estimate for \Nga is then given by
\begin{equation}
    \Nga = \frac{\Delta \gobsa}{\Delta \gtruea}\label{eq:Nga} ,
\end{equation}
where $\Delta \gobsa$ is the change in measured mean shear w.r.t a simulation with constant shear $\gconst$ at all redshifts.

In practice, we always use $\Delta \gtruea = 0.04$ and $\gconst=-0.02$. The latter is chosen because we already generated simulations with constant shear $-0.02$ (separately for both shear components) for the constant shear results in \sect{sec:constant_shear}. That is, for each redshift interval $\alpha$, we generate an extra simulation with 
\begin{align}
(g_1,g_2) =
\begin{cases}
(0.02,0.00) & \text{if $z_1^{\alpha}<z<z_2^{\alpha}$} \\
(-0.02,0.00) & \text{otherwise}.
\end{cases}
\end{align}
This allows us to estimate \Nga\ for the first shear component; we also generate an analogous simulation to estimate it for the second shear component, with 
\begin{align}
(g_1,g_2) =
\begin{cases}
(0.00,0.02) & \text{if $z_1^{\alpha}<z<z_2^{\alpha}$} \\
(0.00,-0.02) & \text{otherwise},
\end{cases}
\end{align}
and then average over the two sets of per-shear-component \Nga\ measurements. 

We use four redshift intervals $\alpha$ with lower and upper redshift limits $(z_1^\alpha,z_2^\alpha)\in \{(0.0,0.4), (0.4,0.7), (0.7,1.0), (1.0,3.0)\}$. The bottom section of \tab{tab:simulations} summarizes the simulation inputs for these redshift-dependent shear simulations.

Since we have only a finite volume of simulation and thus noisy estimates of $\Delta\gobsa$, and the estimates for different $\alpha$ are correlated (for two reasons - they are differences with a common constant shear simulation, and their random seeds are matched with that constant shear simulation to reduce noise on the differences), we construct the covariance matrix $\text{Cov}(\Delta\gobsa, \Delta g^{\text{obs}}_\beta$) by jackknifing over simulated tiles. We then have
\begin{equation}
    \text{Cov}(\Nga,N_\gamma^\beta) = \text{Cov}(\Delta\gobsa, \Delta g^{\text{obs}}_\beta) / (0.04^2).
\end{equation}

\subsection{Effective redshift distribution for lensing: Simulation measurements vs. predictions}
\label{sec:Ngamma_nontomo}

We start with the simplest case of \Nga\ for the full ensemble of detected objects in our simulations which pass the standard selection cuts - we call this the `non-tomographic` case, since no tomographic redshift binning is applied. 
As discussed in \sect{sec:shear_biases}, we are interested foremost in measuring biases in the estimation of $n_{\gamma}(z)$ via the method used on the real data - that is, a \mcal\ response-weighted histogram of redshift estimates, which we denote $n_{\gamma}^{\text{mcal}}(z)$. From this, one can make a prediction for \Nga\
\begin{equation}
    \Ngamcal = \int_{z_1^\alpha}^{z_2^\alpha}\dx{z} \neffzmcal.
\end{equation}
Since in the simulations we work with discrete detections with assigned redshifts, we can estimate $\Ngamcal$ via summing the \mcal\ responses of detections assigned to redshift interval $\alpha$:
\begin{equation}
   \Ngamcal  = \frac{\sum_{z_1^\alpha<z_j<z_2^\alpha} R_j}{\sum_j R_j}\label{eq:Ngamcal},
\end{equation}
where the sum in the numerator is over all detected objects with an assigned redshift in the interval $z_1^\alpha<z_j<z_2^\alpha$, and the sum in the denominator is over all detected objects. \Ngamcal\ then is simply the fraction of the response-weighted detections in redshift interval $\alpha$. Any differences between this fraction, and the \Nga\ measured directly from the simulations via \eqn{eq:Nga} would imply a bias in \neffzmcal\ as an estimate of \neffz.


\begin{figure}
\includegraphics[width=0.9\columnwidth, trim=0.5cm 0.5cm 0.5cm 0, clip=True]{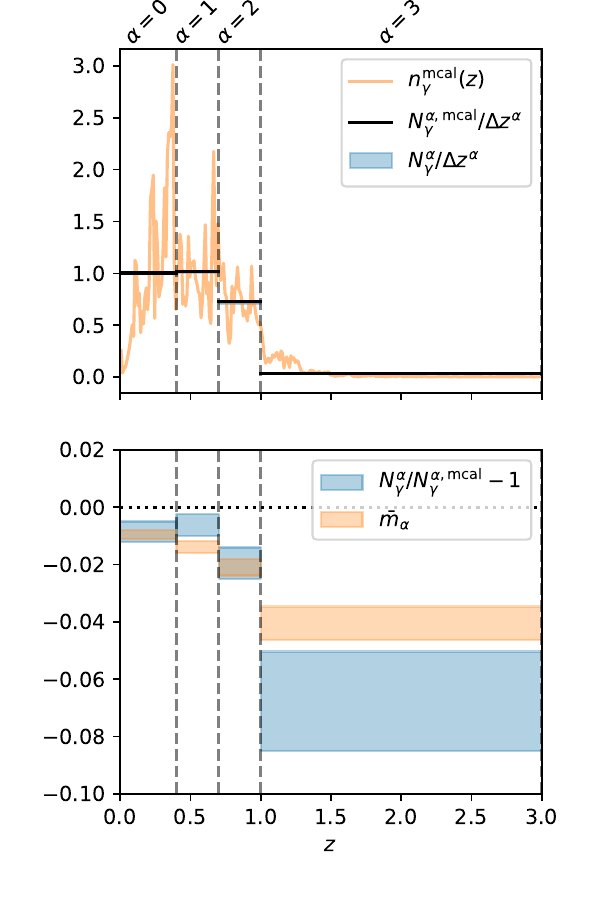}
\caption[]{Measurements of \Nga without \photoz\ binning. Top panel: The orange line is $\neffzmcal$ (the \mcal\ response-weighted $n(z)$), and the black horizontal lines show $\Ngamcal$ i.e. the prediction for $\Nga$ based on integrating $\neffzmcal$ over each interval $\alpha$. We also show as blue rectanges the direct measurements of $\Nga$ from the simulations, with the height of the rectangle indicating the $1\sigma$ uncertainty region. In both cases we divide by the width of the redshift interval $\alpha$ for consistency with the plotted $\neffzmcal$.
On this scale the percent level differences between $\Nga$ (black lines) and $\Ngamcal$ (blue rectangles) are difficult to perceive, hence we show fractional difference in the bottom panel. 

Bottom panel: Blue rectangles indicate the fractional difference between the measured $N_{\gamma}^\alpha$, and $\Ngamcal$. $\Ngamcal$ is biased high, especially for $z>1$, where the bias is around $6\%$. This means that the response of the shear catalog to shear at $z>1$ is around $6\%$ less than predicted. 
Orange rectangles show, $\bar{m}_\alpha$ the mean multiplicative bias for objects assigned to each redshift interval, as inferred from constant shear simulations. The fact that the two sets of measurements differ implies the presence of cross-redshift blending - where galaxies assigned to one redshift interval have some non-zero response to shear applied to a different interval.}
\label{fig:N_gamma_nontomo}
\end{figure}

In the top panel of \fig{fig:N_gamma_nontomo}, we show the simulation measurements of \Nga, as well the \Ngamcal\, and for reference the redshift distribution \neffzmcal. On this scale, the simulation measurements, \Nga, are indistinguishable from the prediction, \Ngamcal, so in the bottom panel we show the fractional difference between the two quantities.
The blue rectangles are the fractional difference between \Nga\ measured from the simulations (\eqn{eq:Nga}) and, \Ngamcal\ (estimated via \eqn{eq:Ngamcal}) i.e. $\Nga/\Ngamcal-1$ for the four redshift intervals $\alpha$.
The height of the rectangles denotes the $1-\sigma$ error bounds on this fractional difference, propagated from the covariance on the measured \Nga. We see that \Ngamcal is a percent-level overestimate of \Nga, with the disparity  increasing for the higher redshift intervals. 

What is going on here? Let us consider the highest redshift interval. \fig{fig:N_gamma_nontomo} implies that our ensemble of detections responds to a shear in this redshift interval around $6\%$ more weakly than one would expect from the \mcal\ response-weighted redshift distribution. This could be an indication of either of two subtly different effects (or both):
\begin{enumerate}
    \item{The measured shear for detections assigned redshift $z$ depends only on the applied shear at redshift $z$, but has some (redshift-dependent) mean multiplicative bias $\bar{m}(z)\neq0$ or equivalently non-unity mean response $\bar{R}(z)\neq1$ (that is not captured by \mcal), such that $\bar{\gamma}^{\textrm{obs}}(z) = \bar{R}(z)\gamma^{\textrm{true}}(z)$, and therefore $\neffz=\bar{R}(z)\neffzmcal$. }
    \item{Due to blending, the measured shear for detections assigned redshift $z$ has some additional non-zero response to the applied shear at other redshifts $z'$ (due to contamination by light from galaxies at $z'$), such that $\bar{\gamma}^{\textrm{obs}}(z) = \int R(z,z') \gamma^{\textrm{true}}(z')$. Here $R(z,z')$ is some function that determines the strength of linear response to shear applied at redshift $z'$, for detections assigned redshift $z$. We might expect this sort of effect from blending, since the measured shear of objects assigned redshift $z$ may be influenced by the shear applied to light from other redshifts $z'$. This is precisely the effect we demonstrated with our simple simulation in \sect{sec:intro}, where the shape measurement of the high $z$ galaxy had non-zero response to shear applied to the low $z$ galaxy.}
\end{enumerate}

We know to some extent the first effect is present - we do see a multiplicative bias in the constant shear simulations in \sect{sec:constant_shear}.
In the next section we show measurements of $N_{\gamma,\alpha}^i$, for true redshift bins $i$, which gives us a clear insight into whether the second effect is present. But before that, is useful to consider $\bar{m}_{\alpha}$, the multiplicative bias for detections assigned to redshift interval $\alpha$ as measured from constant shear simulations. If only the first mechanism above was present, we would expect this measurement to be equivalent to the fractional difference $\Nga/\Ngamcal-1$ shown in \fig{fig:N_gamma_nontomo}. The orange rectangles in \fig{fig:N_gamma_nontomo} show $\bar{m}_{\alpha}$, and the difference with respect to $\Nga/\Ngamcal-1$ implies the presence of the second mechanism above, as we will explore further in the next section.

\subsection{$N_{\gamma}^{\alpha}$ measurements for assigned true redshift bins}\label{sec:Ngamma_slice}

\begin{figure}
\includegraphics[width=0.9\columnwidth]{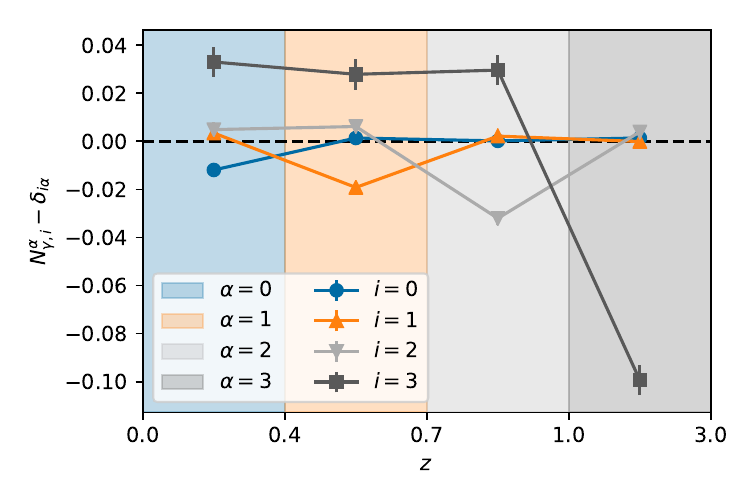}
\caption[]{Measurements of $N_{\gamma,i}^{\alpha}$ for \emph{assigned true} redshift interval $i$. $N_{\gamma,i}^{\alpha}$ is the response of the mean shear of galaxy ensemble $i$, to a shear applied in redshift interval $\alpha$. In this case our ensembles $i$ correspond to intervals in true redshift. 
The \mcal\ prediction for $N_{\gamma,i}^{\alpha}$ in this case is simply the identity matrix: $N_{\gamma,i}^{\alpha}=\delta_{i\alpha}$, since one would expect unity response of galaxies to shear applied in their own redshift interval, and zero response otherwise. Hence we have subtracted $\delta_{i\alpha}$ in this plot to reduce the dynamic range. The $y$-axis then demonstrates the biases  from assuming the \mcal-response weighted $n(z)$ as the effective redshift distribution.}
\label{fig:N_gammaalphai_nontomo}
\end{figure}

We noted in \sect{sec:ngamma2} that it will be useful to measure $N_\gamma^\alpha$ for subsets $i$ of our detections. While using photometric redshift bins as those subsets, as we do in the next section, is more directly applicable to an analysis of real data, it is instructive to study $N_{\gamma,i}^\alpha$ for bins $i$ in ``true'' redshift, or really \emph{assigned} true redshift. The qualification here is important because we have already noted that \emph{detections} do not necessarily correspond to only one galaxy, so may not have a unique true redshift, but they are assigned a unique true redshift in the matching procedure described in \sect{sec:sompz}. 

A measurement of $N_{\gamma,i}^\alpha$ for $i \neq \alpha$ directly probes the presence of effect (ii) discussed above -- the linear response of the shear measured for detections assigned a redshift $z$, to the applied shear at other redshifts $z' \neq z$.  It is calculated via 
\begin{equation}
    N_{\gamma,i}^\alpha = \frac{\Delta\bar{\gamma}^{\textrm{obs}}_{\alpha,i}}
    {\Delta\gamma^{\textrm{true}}_\alpha},
\end{equation}
where $\Delta\bar{\gamma}^{\textrm{obs}}_{\alpha,i}$ is the change in mean measured shear for galaxies assigned to redshift interval $i$ only, for the simulation in which redshift interval $\alpha$ is sheared. $N_{\gamma,i}^\alpha$ can be related to the $R(z,z')$ postulated in \sect{sec:Ngamma_nontomo} via
\begin{equation}
    N_{\gamma,i}^\alpha = \int_{z_1^i}^{z_2^i} dz \int_{z_1^\alpha}^{z_2^\alpha} dz' R(z,z').
\end{equation}

The prediction for $N_{\gamma,i}^\alpha$ based on the \mcal\ response-weighted $n(z)$ for detections assigned to redshift interval $i$, \neffizmcal, is simple. Since in this case the full sample is within redshift interval $i$, we should see unity response to shear applied to the assigned interval $i$ i.e. when $\alpha=i$, and zero response otherwise i.e. when $\alpha \neq i$. That is, $N_{\gamma,i}^{\text{mcal}}=\delta_{i,\alpha}$.

In \fig{fig:N_gammaalphai_nontomo}, we plot as points with errorbars the $N_{\gamma,i}^\alpha$ for the four redshift intervals used. Since we expect $N_{\gamma,i}^\alpha$ to be close to unity for $i=\alpha$, we subtract $\delta_{i\alpha}$ from the measurements so that the results are all near zero. We see that the diagonal elements of \Ngia are all less than unity, implying that detections assigned to a given redshift interval do not have unity response to shear applied in that same redshift interval, with the size of the effect increasing from around 1\% at low redshift to around 10\% at high redshift. This implies the presence of a multiplicative shear bias e.g. due to a dilution of the applied shear due to contamination of the shape measurement by light from other (unsheared) redshift intervals. 

The off-diagonal terms in $N_{\gamma,i}^\alpha$ are predominantly positive, especially for $i=3$, that is for detections assigned to the highest redshift interval. This means that the detections exhibit a positive response to shear applied in the other redshift intervals $\alpha \neq i$, a clear detection of the mechanism (ii) described in the previous section, and demonstrated in our simple simulation in \sect{sec:intro}.
This effect is potentially important since it implies there will be extra correlation in the shears measured at different redshifts w.r.t what one would expect from the estimated redshift distribution $\neffzmcal$.  Or put differently, there will be additional tails or broadening in the effective redshift distribution for weak lensing.

\subsection{$N_{\gamma}^{\alpha}$ measurements for \photoz\ bins}\label{sec:Ngamma_tomo}

We now repeat the above measurement, but applied to photometric redshift bins, rather than bins of assigned true redshifts.
These are the most directly applicable measurements for the DES Y3  weak lensing analyses. 
We use the same simulations and shear measurements to
calculate the integrated effective density \Ngia\ for photometric redshift bin $i$.  \Ngia\ is estimated from the simulations in the same way as above but now simply restricting the mean shear calculation to detections assigned to photometric redshift bin $i$, according to the procedure described in \sect{sec:sompz}. 
Note the change in meaning of index $i$, which now labels the \photoz\ bin rather than the bin in assigned true redshift. 

\begin{figure*}
\includegraphics[width=0.9\textwidth]{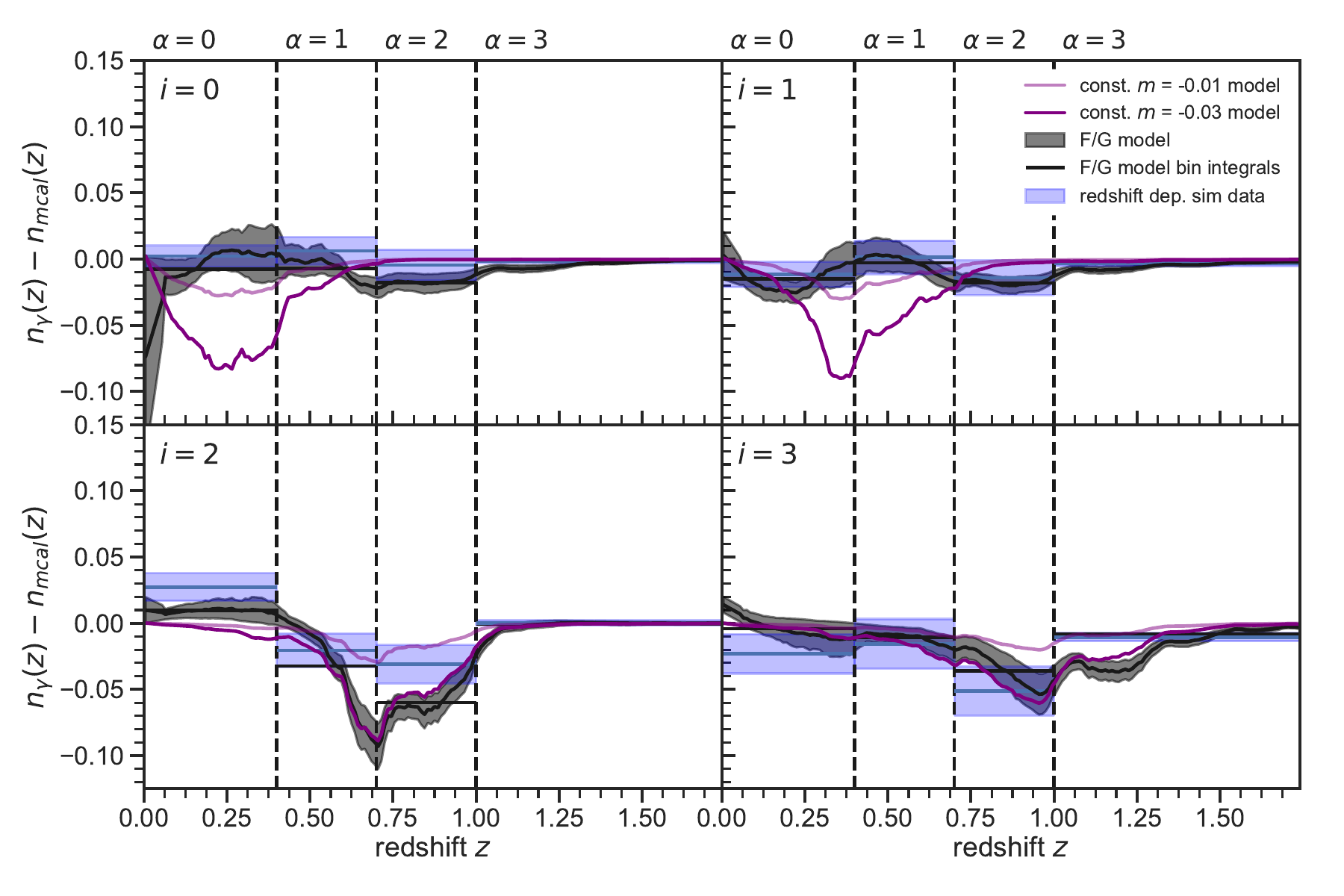}
\caption[]{
  Modeling \neffizmodel\ using \Ngia\ from our image simulations. In each panel, the blue rectangles show $\Ngia-\Ngamcal$ for the photometric redshift bin $i$. The width represents the redshift extent of the interval $\alpha$ and height indicates the $1\sigma$ uncertainty on the measurement. The grey band in each panel shows the posteriors of our \neffizmodel\ and the horizontal black lines are the integrals over this model which form the predictions for the \Ngia\ measurements from the simulations. The light and dark purple lines show the models that would result if one used a constant multiplicative bias of $-0.01$ and $-0.03$ respectively. It is clear from the simulation data, especially in bin 2, that no constant multiplicative bias model would be sufficient. In particular, at lower redshift in bin 2, they cannot capture the increased response to low redshift shear due to blending.
 } 
\label{fig:Ngia_tomo}
\end{figure*}

\fig{fig:Ngia_tomo} shows the difference between the \Ngia\ measurements and the \Ngiamcal\ predictions for each \photoz\ bin $i$. 
In this case, \Ngiamcal\ is no longer $\delta_{i,\alpha}$ as before, since for any $(i,\alpha)$ pair, the \photoz\ bin $i$ includes galaxies outside of the range $(z_1^\alpha,z_2^\alpha)$.
Instead, we now calculate \Ngiamcal\ using \eqn{eq:Ngamcal}, summing over the \mcal\ responses for the galaxies in each \photoz\ bin $i$.
The blue rectangles in \fig{fig:Ngia_tomo} show the resulting $\Ngia - \Ngiamcal$ values with their estimated uncertainties.

The measurements are noisier for this case, especially where there is little weight in the underlying $n(z)$,
e.g. at high redshift ($\alpha=2$ and $\alpha=3$) for the first \photoz\ bin, $i=0$. Due to the fact that our photometric redshift pipeline cannot group galaxies perfectly in redshift, the effects from source blending in Figure~\ref{fig:N_gammaalphai_nontomo} are diluted. The model fits that also appear in Figure~\ref{fig:Ngia_tomo} are described in the next section.

\subsection{$n_{\gamma}(z)$ bias model constraints for \photoz\ bins}\label{sec:ngammamodel}

Above, we have made measurements of $\Ngia$ in a set of coarse redshift bins. While these measurements constitute direct 
constraints on the shear systematics in these bins, we expect the underlying effects to be smooth functions of redshift, and we require a continuous corrected \neffz\ to make theoretical predictions. Thus we now fit a smooth 
model to the measurements. We posit that this smooth model has two terms as follows
\begin{equation}\label{eq:ngamma_model}
    \neffizmodel = \left(1+F_i(z)\right)\neffizmcal + G_i(z) \; .
\end{equation}
The term $F_i(z)$ captures changes in the effective weighting of the discrete sources used to construct $\neffizmcal$. These 
effects could include things like constant multiplicative biases or redshift-dependent multiplicative bias effects. The term 
$G_i(z)$ captures responses to shear at redshifts not present in the naive $\neffizmcal$ distribution computed from the detected 
sources. This term is designed to capture blending effects where high-redshift sources respond to input shear at lower redshifts. Note though that a sufficiently flexible $F_i(z)$ could also capture such effects.

\subsubsection{The model}

Due to the limited volume of image simulation data, we can only employ relatively constrained models. For $G_i(z)$, we fit a template 
with a single free amplitude. The template is built from the flux density of the COSMOS sample in the $riz$-bands as a function of redshift.
This model is motivated by the idea that for random projections, blending effects will be proportional to the flux density of objects.

For $F_i(z)$, we employ a model with two parts. First, we have a constant term that is meant to capture overall multiplicative biases. 
Second, we employ a perturbative method detailed in \citet{gramchar} to build a set of correction terms that capture mean shifts 
in the redshift distribution and possibly changes in its width. The overall idea of this method is that we can expand \neffizmodel\ perturbatively 
about the input \neffizmcal\ with the coefficients of the expansion corresponding to changes in the moments of \neffizmcal. The form of this series 
for two general PDFs $h(x)$ and $p(x)$ is
\begin{equation}
h(x) \approx p(x)\left(1 + a_1\frac{1}{p(x)}\frac{dp(x)}{dx} + a_2\frac{1}{2p(x)}\frac{d^2p(x)}{dx^2} + \ldots\ \right).
\end{equation}
In our application, we use either the first or the first two terms of this series to generate template functions to put into $F_i(z)$. We then fit 
for the coefficients $a_1$ and/or $a_2$. Given that directly differentiating the simulation \neffizmcal\ will be quite noisy, we fit a model to this quantity 
and then differentiate that model. If we call this model $\phi(z;\theta_i)$, then the final form for $F_i(z)$ is 
\begin{equation}\label{eq:fmodel}
    F_i(z) = a_0 + a_1 \frac{1}{\phi(z;\theta_i)}\frac{d\phi(z;\theta_i)}{dz} + a_2 \frac{1}{2\phi(z;\theta_i)}\frac{d^2\phi(z;\theta_i)}{dz^2}\; .
\end{equation}
In this expression, $a_0$ is approximately an overall multiplicative bias term, $a_1$ approximately controls changes in the mean redshift of \neffizmcal, 
and $a_2$ approximately controls a combination of the mean and the width of \neffizmcal. We use the following distribution \citep{baugh1993,brainerd1996},
\begin{equation}\label{eq:fidnz}
\phi(z;a,b,c) \propto z^a \exp\left(-\left(\frac{z}{b}\right)^c\right)
\end{equation}
for our fiducial model for $\phi$, keeping only the terms up to $a_1$ in \eqn{eq:fmodel}. As an alternative, we also considered
using a Student's-$t$ distribution with $\nu = 1$ degree of freedom, keeping all terms up to $a_2$. We found that for the Student's-$t$ distribution we needed to increase the width of the distribution by a factor of 10 when generating the model terms in order to avoid extremely strong model effects in the tails of the distribution.

From these smooth models, we can estimate \Ngia\ via 

\begin{equation}
    \Ngiamodel = \int_{z_1^\alpha}^{z_2^\alpha} \dx{z} \neffizmodel.
\end{equation}
We can also make a prediction for the mean multiplicative bias for \photoz\ bin $i$,  $\bar{m}_i$, or equivalently mean response $\bar{R}_i=1+\bar{m}_i$, which we measured from the constant shear sims, which is simply
\begin{equation}
    \bar{R}_i^{\text{model}} = \int_0^\infty \neffizmodel ,
\end{equation}
i.e. the normalization of $\neffz^{\text{model}}$. Therefore we can use both $\Ngia$ and $\bar{R}_i$ measurements to constrain the $F_i(z)$ and $G_i(z)$ components of our $\neffz$ bias model.

\subsubsection{Fitting procedure}\label{sec:fgfitting}

Equipped with simulation measurements of $\Ngia$ and $\bar{R}_i$, 
an estimate of their joint covariance matrix $C_i$ (again estimated by jackknifing over simulated tiles)
and a model for $\neffz$ with which we can predict them, we can now form a likelihood to constrain that model.

We use Markov chain Monte Carlo (MCMC) to produce samples of $F(z)$ and $G(z)$ and thus $n_{\gamma}^{\text{model}}(z)$. Our likelihood is
\begin{equation}
    \log L = \mathcal{N}\left(\left[\bar{R}_i, \Ngia \right]^{\text{sim}} - 
    \left[\bar{R}_i, \Ngia \right]^{\text{model}}, C_i\right),\label{eqn:like}
\end{equation}
where $\mathcal{N}(\mu,\Sigma)$ indicates a Gaussian distribution with mean $\mu$ and covariance $\Sigma$. We use wide, uninformative priors on the free parameters of $F(z)$ and $G(z)$, and use \textsc{emcee} \citep{emcee} to sample the posterior. We use a configuration of 12 walkers taking $10^4$ steps each. We then discard the first $5\times10^3$ samples and thin the rest of the samples by a factor of 10. Our final chains have autocorrelation lengths of approximately $40-50$ and are thus long enough to generate a sufficient number of independent samples. Having sampled the posteriors of the model parameters, we can then produce samples of $F(z)$, $G(z)$, and $\neffizmodel$ that are conditioned on our simulation measurements. 

\subsection{Model constraints}\label{sec:model_constraints}

The posterior constraints on our models are show in \fig{fig:Ngia_tomo} as the grey bands. We also show the predictions for \Ngia\ as the horizontal grey lines. We find that our model generally provides a good fit to the data. In this figure we also show the predictions of a model which employs a constant $m$ per tomographic bin, i.e. $\neffizmodel = (1+m_i)\neffizmcal$. While this model may be sufficient for the lower redshift tomographic bins, in bin $i=2$ we find that it cannot capture the low redshift tails introduced into $\neffizmodel$, perhaps due to blending. 

From each $\neffizmodel$ sample we can calculate an effective multiplicative bias $m$, and a change in mean redshift \Dz.
$m$ is equal to the normalization of $\neffizmodel$ minus 1, i.e.
\begin{equation}\label{eq:m_from_ngamma}
m = \int \dx{z} \neffizmodel - 1
\end{equation}
and 
\begin{equation}\label{eq:dz_from_ngamma}
    \Dz = \frac{\int \dx{z} z \neffizmodel}{\int \dx{z} \neffizmodel} - \frac{\int \dx{z} z \neffizmcal}{\int \dx{z} \neffizmcal}.
\end{equation}
Note we have chosen the sign convention of our definition of $\Dz$ to match that used in the cosmology analyses (eg. \citet{y3-3x2ptkp}), which means it is the correction one would apply to \neffizmcal, rather than the bias in \neffizmcal\ (which would have the opposite sign).

\begin{table}
\begin{tabular}[width=0.9*\textwidth]{c|c|c|c}
\hline
quantity & tomo. bin & fiducial model & Student's-$t$ model \\ 
\hline
$m\times100$               & 0 & $-1.36 \pm 0.29$ & $-1.37 \pm 0.29$ \\
$m\times100$               & 1 & $-1.78 \pm 0.36$ & $-1.77 \pm 0.36$ \\
$m\times100$               & 2 & $-2.48 \pm 0.41$ & $-2.52 \pm 0.42$ \\
$m\times100$               & 3 & $-3.34 \pm 0.52$ & $-3.31 \pm 0.52$ \\
\hline
$\Delta\bar{z}_0\times100$ & 0 & $-0.65 \pm 0.27$ & $-0.96 \pm 0.30$ \\
$\Delta\bar{z}_0\times100$ & 1 & $-0.80 \pm 0.31$ & $-0.76 \pm 0.29$ \\
$\Delta\bar{z}_0\times100$ & 2 & $-0.36 \pm 0.26$ & $-0.40 \pm 0.19$ \\
$\Delta\bar{z}_0\times100$ & 3 & $-1.13 \pm 0.60$ & $-0.45 \pm 0.37$ \\
\hline
\end{tabular}
\caption{\label{tab:mdz} Inferred constraints on $m$ and $\dz$ from the \neffz\ model samples fit to the image simulations measurements. These \neffz\ models are generated by perturbing the image simulation \neffzmcal.}
\end{table}

These two summary statistics of \neffz\ are probably the two most important for weak lensing analysis, where higher-order biases in the redshift distribution are likely to have a subdominant effect to changes in normalization (i.e. multiplicative bias) or mean redshift.
The two models give largely similar constraints in $m$ and $\dz$, with the fiducial model in general allowing slightly larger shifts in redshift. The first two lines of \tab{tab:mdz} contains these inferred $m$ and $\dz$ constraints for the two \neffz\ bias models (fiducial and Student's-$t$), when applied to the fiducial $n(z)$s from the image simulations. 

The \neffz\ bias model constraints encapsulate our characterization of shear calibration biases for each DES Y3 photometric redshift bin, including redshift dependent effects such as the impact of blending on the effective redshift distribution for lensing. They also include the statistical uncertainty on the calibration due to finite simulation volume. When using the \neffz\ bias model for the real DES Y3 data, we apply it instead to the $n(z)$s estimated from the DES Y3 data. In the next section, we describe this procedure, and explore potential systematic uncertainties in our corrections, in order to provide statistical and systematic uncertainty priors on the corrections measured from the simulations here.

\section{\neffz\ corrections for the DES Y3 shear catalog}\label{sec:y3nz}

In \sect{sec:zslice_shear}, we generated estimates of the bias in our methodology for constructing the effective redshift distribution for lensing, \neffz. We did this by comparing direct estimates of \neffz\ from the image simulations, possible only because we know the input shear to the simulations, to the method we use on the Y3 data: the \mcal\ response-weighted redshift distribution, \neffzmcal. These bias estimates are intended to be used to make corrections to the redshift distributions used in theory predictions in weak lensing cosmology analyses. In our case, those redshift distributions are those estimated from the DES Y3 data, summarized in \citet*{y3-sompz}, which provide ensembles of redshift distributions constrained by both galaxy flux information, and clustering \citep*{y3-sourcewz}. In \sect{sec:apply_to_data} we describe our procedure for applying our corrections to these redshift distributions. This procedure results in a set of \neffz\ samples, which can be sampled over when performing cosmological parameter inference (e.g. using the \textsc{hyperrank} method described in \citealt{y3-hyperrank}).

The accuracy of our corrections depends on how realistically we have simulated the DES Y3 data. In Sections \ref{sec:m_pzbinning}-\ref{sec:m_fgmodel} we explore this issue by inspecting the sensitivity of our \neffz\ bias corrections to variations in the simulation measurements or simulation analysis choices. Results from these variations feed into our systematic uncertainties.

Finally, in \sect{sec:final_prior} we summarize our final corrections which include both statistical and systematic uncertainties.

\subsection{Applying corrections to the DES Y3 $n(z)$s}\label{sec:apply_to_data}

\citet*{y3-sompz} provide an ensemble of 1000 samples (for each of the four \photoz\  bins) drawn from the posterior of the set of \photoz\ bin redshift distributions, constrained by both galaxy flux (including color) and clustering information. These distributions are already weighted by the \mcal\ response estimated from the DES data, so in fact correspond to samples of \neffizmcal.
For each of these 1000 samples, we draw 100 randomly chosen samples of our bias model inferred from our simulation measurements, quantified by the functions $F(z)$ and $G(z)$.  
Then we apply these model biases to \neffizmcal\ using \eqn{eq:ngamma_model} to produce estimates of $\neffz$.  This process leaves us with $1000\times100=100,000$ \neffz\ samples for each \photoz\ bin, but note that in practice one could downsample this set of $\neffz$s to a more manageable number depending on the application.

\begin{figure*}
\includegraphics[width=0.9\linewidth]{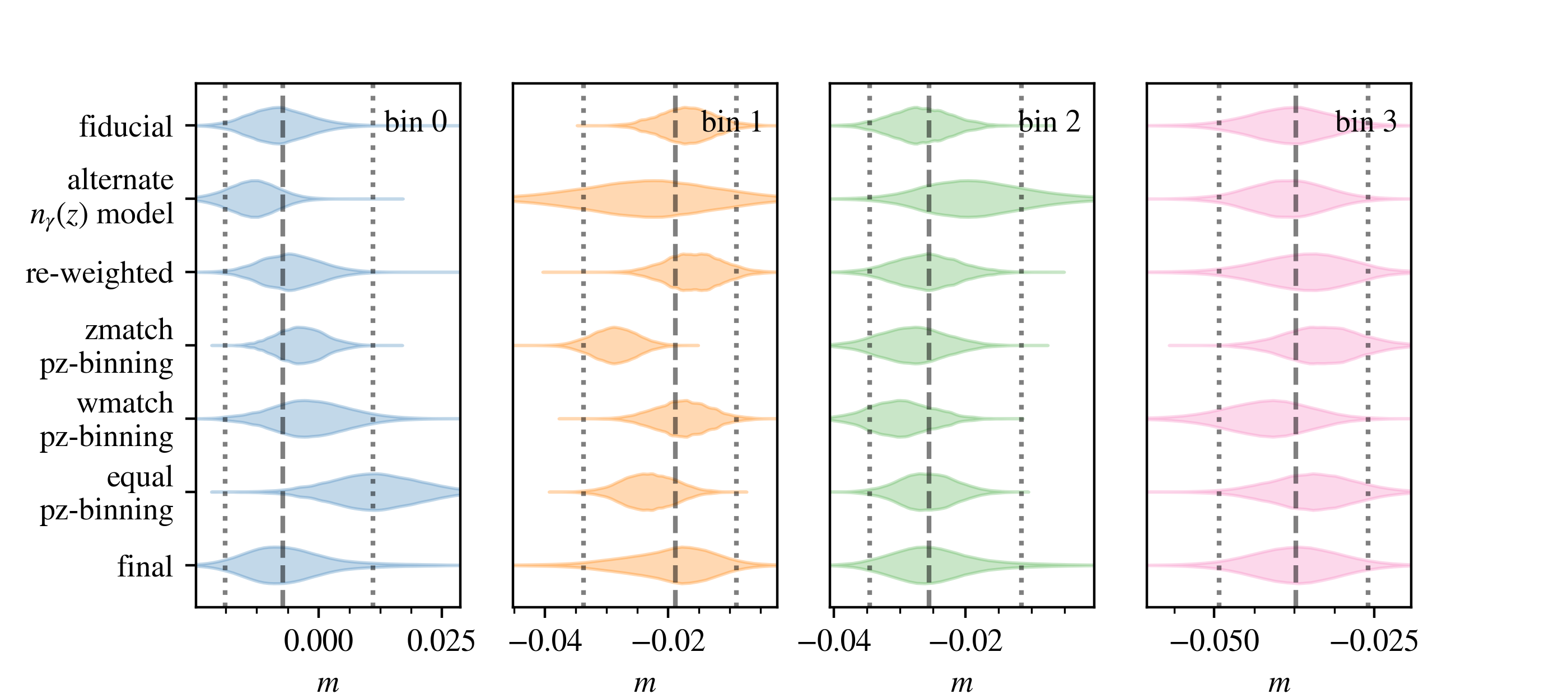}
\includegraphics[width=0.9\linewidth]{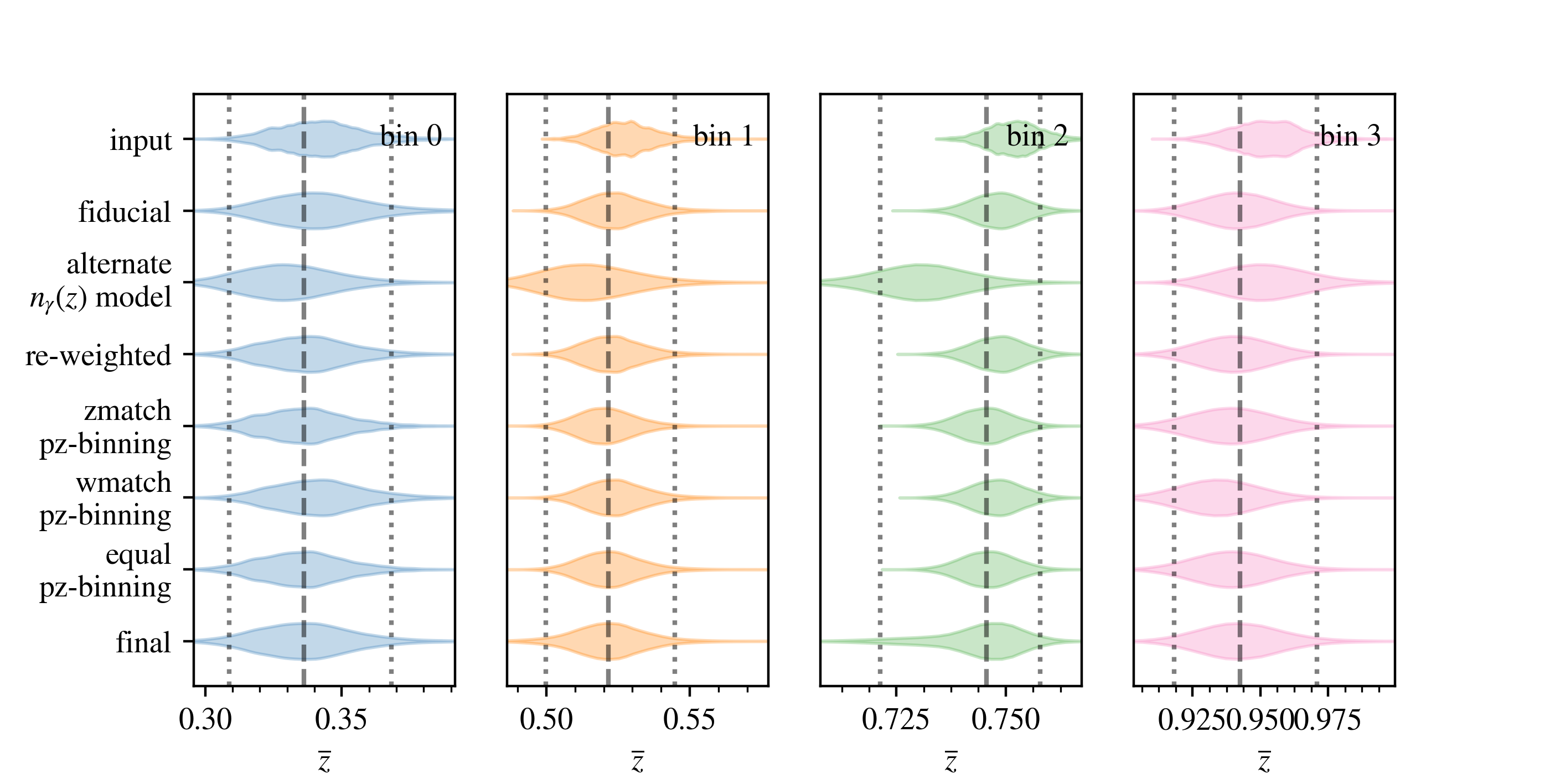}
\caption[]{
    Distributions of the inferred $m$ (top panels) and mean redshift $\bar{z}$ (bottom panels) of our  DES Y3 redshift \neffz\ estimates, generated by applying our simulation corrections to an input ensemble of redshift distributions.
    For each line \neffz\ corrections are applied from a variation on the fiducial simulation analysis, which have been chosen to expose potential systematic errors in the fiducial approach. The lines marked `final' represent the mixture approach described in \sect{sec:final_prior}, which incorporates systematic uncertainty by mixing together samples from the alternative approaches with the fiducial analysis.
    In the lower panel, we show also in the first line the distribution of mean redshifts for the input (i.e. before correction) redshift distributions. 
}
\label{fig:violin}
\end{figure*}

In \fig{fig:violin} we show the distributions of mean multiplicative bias, $m$ (top panels), and mean redshift, $\bar{z}$ (bottom panels) inferred from these \neffz\ samples. $m$ is calculated as in \eqn{eq:m_from_ngamma} i.e. is simply the normalization of \neffz\ minus 1. In all cases the height of the violin shape is proportional to the fractional weight in the distribution at a given value of the $x$-axis.
The lines labelled "fiducial" represent the inferred $m$ and $\bar{z}$ distributions for this case. We also list the mean and standard deviation of these $m$ and $\bar{z}$ distributions in \tab{tab:priors}. We note here that we are in some sense extrapolating the corrections we have inferred from our simulations onto the DES Y3 data, and one must always be careful with extrapolation. However, the fact that the mean $m$ values inferred from the image simulation \neffz\ models (first row of \tab{tab:mdz}) are within the $1\sigma$ bounds of the "fiducial" Y3 data $m$ priors in \tab{tab:priors}, gives us confidence that this extrapolation is well-behaved.

In \tab{tab:priors} we also list the mean and standard deviation of two other quantities, labelled ``\dz'' and ``$\dz_{FG}$''. These are defined as follows. For a given input \neffzmcal\ sample, which we will call $n^{\text{mcal}}_s(z)$, we generate an output $\neffz$ sample, $n_{\gamma,s}(z)$, by applying samples of the $F(z)$ and $G(z)$ perturbation functions. For a given $n_{\gamma,s}(z)$, $\dz_s$ is defined as the change in mean redshift with respect to the mean redshift of the \emph{ensemble} of input \neffzmcal\ samples i.e.
\begin{equation}
    \dz_s = \bar{z}\left[n_{\gamma,s}(z)\right] - \left<\bar{z}\left[n^{\text{mcal}}_s(z))\right]\right>_s ,
\end{equation}
where $\bar{z}\left[\psi(z)\right]$ represents the mean redshift of the function $\psi(z)$, and $\left<\right>_s$ represents an average over all samples $s$. Note the distribution of this \dz\ will include the scatter in the input \neffzmcal\ samples. 

The quantity $\dz_{FG}$ on the other hand does not; it is defined as
\begin{equation}
    \dz_{FG,s} = \bar{z}\left[n_{\gamma,s}(z)\right] - \bar{z}\left[n^{\text{mcal}}_s(z)\right]
\end{equation}
for a given sample $s$, 
and thus isolates the impact of applying the $F(z)$ and $G(z)$ correction functions. Hence it is this quantity that is more directly comparable to the \Dz\ we defined in \sect{sec:model_constraints} for the image simulation \neffz, and listed values for in \tab{tab:mdz}. The image simulation \Dz\ from \tab{tab:mdz} are all within the $1\sigma$ bounds of the $\dz_{FG}$ distributions, except for \photoz\ bin 1, where it is within $2\sigma$. 

In the next three sub-sections we test for sensitivity of the inferred $m$ and \dz\ priors to various potential systematics in our simulation corrections.

\begin{table*}
\begin{tabular}[width=0.9*\textwidth]{c|c|c|c|c|c }\hline
Model & photo-$z$ bin & $m \times 100$ & $\bar{z}$ & $\dz \times 100$ & $\dz_{FG}\times100$\\
\hline
input & 0 & N/A & $0.341 \pm 0.016$ & $0.000 \pm 1.608$ & $0.000 \pm 0.000$ \\
fiducial & 0 & $-0.789 \pm 0.625$ & $0.341 \pm 0.019$ & $-0.021 \pm 1.938$ & $-0.021 \pm 0.711$ \\
alternate $\neffz$ & 0 & $-1.346 \pm 0.541$ & $0.330 \pm 0.018$ & $-1.054 \pm 1.834$ & $-1.054 \pm 0.991$ \\
re-weighted & 0 & $-0.574 \pm 0.615$ & $0.337 \pm 0.017$ & $-0.406 \pm 1.750$ & $-0.406 \pm 0.353$ \\
zmatch pz & 0 & $-0.368 \pm 0.481$ & $0.335 \pm 0.016$ & $-0.565 \pm 1.644$ & $-0.565 \pm 0.143$ \\
wmatch pz & 0 & $-0.172 \pm 0.790$ & $0.341 \pm 0.018$ & $-0.009 \pm 1.766$ & $-0.009 \pm 0.361$ \\
equal pz & 0 & $1.117 \pm 0.908$ & $0.335 \pm 0.017$ & $-0.580 \pm 1.651$ & $-0.580 \pm 0.212$ \\
\hline
final & 0 & $-0.627 \pm 0.908$ & $0.336 \pm 0.018$ & $-0.467 \pm 1.846$ & $-0.467 \pm 0.758$ \\
\hline
input & 1 & N/A & $0.528 \pm 0.013$ & $-0.000 \pm 1.280$ & $0.000 \pm 0.000$ \\
fiducial & 1 & $-1.703 \pm 0.446$ & $0.524 \pm 0.013$ & $-0.367 \pm 1.344$ & $-0.367 \pm 0.215$ \\
alternate $\neffz$ & 1 & $-2.315 \pm 1.080$ & $0.515 \pm 0.018$ & $-1.267 \pm 1.812$ & $-1.267 \pm 1.441$ \\
re-weighted & 1 & $-1.589 \pm 0.461$ & $0.523 \pm 0.013$ & $-0.467 \pm 1.341$ & $-0.467 \pm 0.193$ \\
zmatch pz & 1 & $-2.869 \pm 0.375$ & $0.521 \pm 0.012$ & $-0.725 \pm 1.212$ & $-0.725 \pm 0.358$ \\
wmatch pz & 1 & $-1.763 \pm 0.452$ & $0.524 \pm 0.013$ & $-0.385 \pm 1.335$ & $-0.385 \pm 0.218$ \\
equal pz & 1 & $-2.308 \pm 0.435$ & $0.522 \pm 0.013$ & $-0.561 \pm 1.311$ & $-0.561 \pm 0.282$ \\
\hline
final & 1 & $-1.984 \pm 0.779$ & $0.521 \pm 0.015$ & $-0.671 \pm 1.506$ & $-0.671 \pm 0.833$ \\
\hline
input & 2 & N/A & $0.750 \pm 0.006$ & $-0.000 \pm 0.629$ & $0.000 \pm 0.000$ \\
fiducial & 2 & $-2.667 \pm 0.458$ & $0.746 \pm 0.006$ & $-0.365 \pm 0.632$ & $-0.365 \pm 0.209$ \\
alternate $\neffz$ & 2 & $-1.649 \pm 0.892$ & $0.729 \pm 0.012$ & $-2.048 \pm 1.173$ & $-2.048 \pm 1.251$ \\
re-weighted & 2 & $-2.570 \pm 0.494$ & $0.746 \pm 0.006$ & $-0.355 \pm 0.641$ & $-0.355 \pm 0.152$ \\
zmatch pz & 2 & $-2.759 \pm 0.511$ & $0.743 \pm 0.006$ & $-0.676 \pm 0.637$ & $-0.676 \pm 0.204$ \\
wmatch pz & 2 & $-2.967 \pm 0.483$ & $0.745 \pm 0.006$ & $-0.411 \pm 0.627$ & $-0.411 \pm 0.202$ \\
equal pz & 2 & $-2.558 \pm 0.430$ & $0.743 \pm 0.006$ & $-0.613 \pm 0.626$ & $-0.613 \pm 0.256$ \\
\hline
final & 2 & $-2.412 \pm 0.760$ & $0.741 \pm 0.011$ & $-0.830 \pm 1.071$ & $-0.830 \pm 0.959$ \\
\hline
input & 3 & N/A & $0.946 \pm 0.015$ & $0.000 \pm 1.524$ & $0.000 \pm 0.000$ \\
fiducial & 3 & $-3.794 \pm 0.718$ & $0.934 \pm 0.013$ & $-1.170 \pm 1.317$ & $-1.170 \pm 1.046$ \\
alternate $\neffz$ & 3 & $-3.780 \pm 0.606$ & $0.945 \pm 0.019$ & $-0.060 \pm 1.913$ & $-0.060 \pm 0.977$ \\
re-weighted & 3 & $-3.574 \pm 0.841$ & $0.933 \pm 0.014$ & $-1.304 \pm 1.378$ & $-1.304 \pm 1.333$ \\
zmatch pz & 3 & $-3.253 \pm 0.589$ & $0.933 \pm 0.017$ & $-1.313 \pm 1.698$ & $-1.313 \pm 0.902$ \\
wmatch pz & 3 & $-4.179 \pm 0.852$ & $0.927 \pm 0.015$ & $-1.875 \pm 1.490$ & $-1.875 \pm 1.575$ \\
equal pz & 3 & $-3.442 \pm 0.690$ & $0.933 \pm 0.015$ & $-1.277 \pm 1.548$ & $-1.277 \pm 1.037$ \\
\hline
final & 3 & $-3.692 \pm 0.761$ & $0.936 \pm 0.017$ & $-1.008 \pm 1.672$ & $-1.008 \pm 1.290$ \\
\hline
\end{tabular}\label{tab:priors}
\caption{Summary statistics of the DES Y3 data \neffz\ distributions, or in the case of the lines labelled ``input'', input ensemble of redshift distributions i.e. before correction. The ``input'' lines hence have no associated multiplicative bias $m$. The column labelled $\bar{z}$ is the mean and standard deviation of the distribution of mean redshift of \neffz. The column labelled $\dz$ is the mean and standard deviation of the distribution of $\dz$, which for a given \neffz, is the difference in mean redshift w.r.t to the mean redshift of the input ensemble of $n(z)$. The column labelled $\dz_{FG}$ is the mean and standard deviation of the distribution of $\dz_{FG}$, which for a given \neffz, is the difference in mean redshift w.r.t the mean redshift of the specific $n(z)$ sample used to generate the \neffz.}
\end{table*}

\subsection{Alternative \photoz-binning schemes}
\label{sec:m_pzbinning}

Given the range of viable different \photoz\ binning schemes presented in \sect{sec:sompz}, we test for the sensitivity of our corrections to this choice, and include a corresponding contribution to our systematic uncertainties related to this choice. For each of the \photoz\ binning schemes, we repeat the inference of \neffz\ corrections performed in \sect{sec:ngammamodel}, but now using the slightly different \Ngia\ and $\bar{R}_i$ measurements resulting from changing the \photoz\ binning scheme. This results in sets of $F(z)$ and $G(z)$ samples for each \photoz\ binning scheme, from which we can generate \neffz\ samples for the DES Y3 data, following the procedure in \sect{sec:apply_to_data}. The inferred $m$s and $\bar{z}$s for these cases are also shown in \fig{fig:violin}, as the lines labelled ``zmatch pz-binning'', ``wmatch pz-binning '' and ``equal pz-binning'' (see \sect{sec:sompz} for the details of the binning methods). 




\subsection{Simulation re-weighting}\label{sec:m_popreweight}

The overall agreement between object properties and density in the simulation and real data is very good, both globally and within each redshift bin. 
To validate that any remaining differences would not significantly modify our inferred shear calibration, we focus on several properties most likely to correlate with the amount of blending, which is our main source of shear calibration bias. 

We implement a re-weighting scheme designed to match the small-scale counts of close pairs in the DES Y3 data, including the magnitude dependence of this behaviour. We believe this should provide a first order estimate of the potential impact of ignoring galaxy clustering in our simulations, which although weak for our wide-in-redshift photometric redshift bins, may have non-negligible impact on the amount of blending. 

We choose three  quantities on which to re-weight: the magnitude (calculated from the mean of the flux in the $r$, $i$ and $z$ bands), the distance to the nearest neighbor, and the magnitude  of the nearest neighbor (again based on the  $r$, $i$ and $z$ bands). The neighbor can be any detected object with non-negative flux (i.e. we do not restrict the neighbor candidates to objects passing shape catalog cuts).  We then aim to produce a set of weights to apply to the simulation results that will improve the match of the joint distributions of these quantities between simulations and data.  

\begin{figure*}
\includegraphics[width=0.45\textwidth]{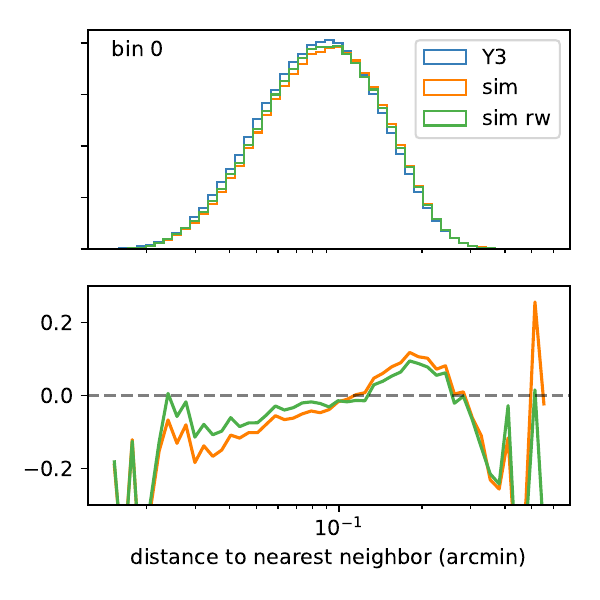}
\includegraphics[width=0.45\textwidth]{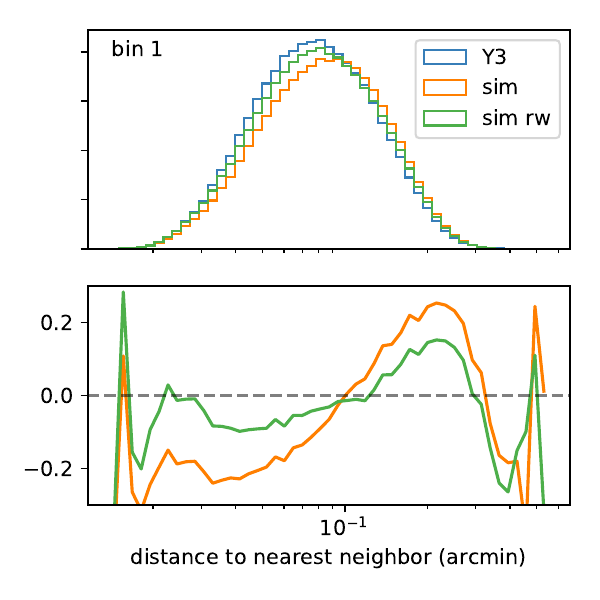}
\includegraphics[width=0.45\textwidth]{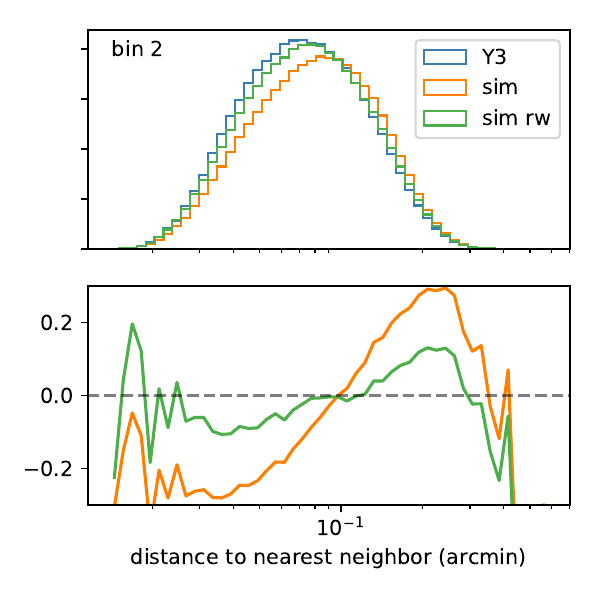}
\includegraphics[width=0.45\textwidth]{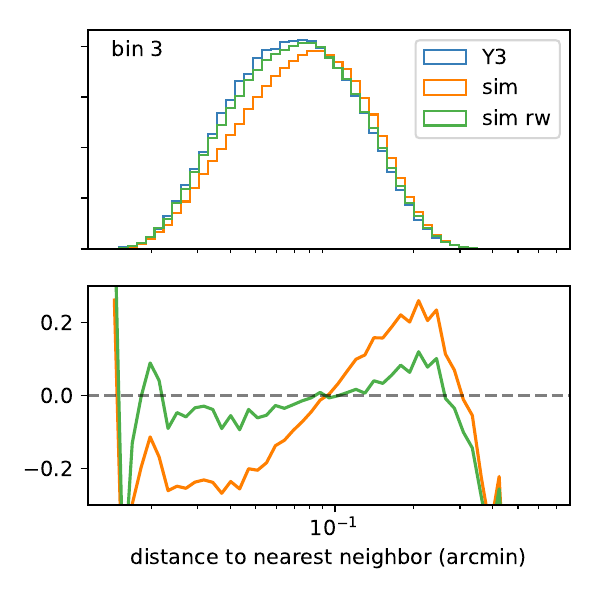}
\caption[]{Distribution of distance to nearest neighbor for each \photoz\ bin. The first and third rows show these distributions (as normalized histograms) for the Y3 data (``Y3'', blue), the image simulations (``sim'', orange) and the re-weighed simulations (``sim rw'', green - using the re-weighting procedure described in \sect{sec:m_popreweight}). The second and fourth rows show the fractional difference of the ``sim'' and ``sim rw'' distributions with respect to the Y3 data distributions. In all cases the agreement between simulations and data is improved by the re-weighting procedure.}
\label{fig:d2n}
\end{figure*}

To accomplish this, we use k-means clustering to define clusters of these three quantities based on 200,000 randomly selected objects in each \photoz\ bin. We then assign all objects in both the simulations and DES Y3 data to these clusters.  Weights are produced for each \photoz\ bin by taking the ratio of the number of the number of objects assigned to each data cluster to the number assigned to each simulation cluster. We show in \fig{fig:d2n} that this re-weighting of objects improves agreement with the data.

We update the weights of our simulated shear catalogs, and re-derive the \neffz\ correction model (by re-performing the  \Ngia\ and $\bar{R}_i$ measurements and the \neffizmodel\ fits). We again apply the resulting $F(z)$ and $G(z)$ corrections to the Y3 data $n(z)$ ensemble to get an ensemble of \neffz\ samples. The inferred $m$s and $\bar{z}$s for these cases are also shown in \fig{fig:violin}, as the lines labelled ``re-weighted''. The impact on $\bar{z}$ is small, with at most 0.2\% shifts in the mean. There is also a small (between 0.1 and 0.2\%), coherent shift in the mean of the $m$ distributions to less negative values for all \photoz\ bins. While small, we find the sign surprising, since we expect this re-weighting procedure to increase the number of close pairs in the simulation, and thus increase the impact of blending on the multiplicative bias. 

We note that this re-weighting procedure will not include all the potentially relevant effects present in a galaxy sample with realistic clustering. We discuss in detail in \sect{sec:conclusions} the possibility that we have underestimated blending-related biases as a result of this. 

\subsection{Alternative \neffz\ bias parameterization}\label{sec:m_fgmodel}

We chose a fiducial functional form for the multiplicative correction $F(z)$ in our \neffz\ bias model. This was based on the perturbative expansion in \eqn{eq:fmodel}, applied to the $n(z)$ model in \eqn{eq:fidnz}. Note that we only use this simple parametric model to generate a functional form for the perturbations from $\neffzmcal$; we do not use the parameteric fit directly.  We have also investigated the use of a Student's-$t$ model, which we use to generate an alternative set of $F(z)$ and $G(z)$ samples with which we correct the DES Y3 $n(z)$ ensemble. The resulting inferred $m$ and $\delta\bar{z}$ distributions are show in \fig{fig:violin}, as the lines labelled ``alternate $n_\gamma(z)$ model''. We observe some non-negligible shifts with respect to the fiducial $m$ and $\bar{z}$. The largest are for \photoz\ bin 2, with a shift in the mean of the $m$ distribution of $1.0\%$, and a shift in the mean $\bar{z}$ of $-1.7\%$. We also see broadening of the $\bar{z}$ distributions for all \photoz\ bins. 


\subsection{Final \neffz\ priors}\label{sec:final_prior}

As discussed in \sect{sec:apply_to_data}, the \neffz\ samples inferred from applying the results of our fiducial simulation to the DES Y3 data $n(z)$ include our fiducial correction and statistical uncertainties. We  incorporate systematic uncertainties as follows. We generate a final prior on \neffz\ which is a mixture of the priors inferred from our fiducial approach, and the variations described in Sections \ref{sec:m_pzbinning}-\ref{sec:m_fgmodel}. More precisely, we formulate the prior $P(\neffz)$ as
\begin{equation}
\begin{split}
P(\neffz) &\propto P(\neffz | \text{Fiducial assumptions})\\ 
+ &P(\neffz | \text{Simulation re-weighting})\\ 
+ &P(\neffz | \text{Alternative \neffz\ bias parameterization})\\ 
+ &P(\neffz | \text{Alternative \photoz\ binning scheme}).
\end{split}
\end{equation}
Each term on the right-hand-side represents the prior on \neffz\ given some set of analysis assumptions. We normalize each term on the right-hand-side such that they contribute equal weight to the total prior on \neffz. Note that this means the priors based on the three alternative \photoz\ binning schemes described in \sect{sec:m_pzbinning} each contribute only one third of the weight to the mixed prior, compared to the other variations. This choice, while somewhat arbitrary, ensures we give equal weight to the three categories of systematic uncertainty defined in Sections \ref{sec:m_pzbinning}-\ref{sec:m_fgmodel}. 

We can generate the final prior on \neffz\ simply by combining the samples from the individual cases in the correct proportions. The inferred $m$ and $\bar{z}$ distributions for this final prior is shown in the bottom rows of \fig{fig:violin}. The grey dotted vertical lines show the 5th and 95th percentiles of this final prior, while the dashed vertical line shows its median. 

One can see qualitatively by eye that this mixture prior accounts for systematic uncertainties with a longer tail where there is an outlier in one of the variations. For example, the $m$ distribution for bin 0, while still centered close to the fiducial case, gains extra weight in the tail to high $m$ due to the outlying $m$ distribution for the "equal" \photoz\ binning variation. Similarly, the significant shift for the alternate $\neffz$ variation in the $\bar{z}$ distribution for bin 2 is accounted for by a large tail to low $\bar{z}$ values.

We note that the $m$ and \dz\ statistics presented here are summary statistics of the full corrections, which are incorporated into perturbed \neffz\ samples. It is these  \neffz\ samples that should be used in theoretical predictions of weak lensing statistics (e.g. shear correlation functions or tangential shear) for the DES Year 3 shear catalog. An algorithm for efficiently sampling over discrete \neffz\ samples, such as that presented in \citet{y3-hyperrank} can be used (or alternatively one could directly sample the likelihood in \eqn{eqn:like}, simultaneously with the  likelihood for the Year 3 weak lensing statistics). However, for some applications, it may be desirable and sufficiently accurate to directly use the inferred $m$ and \dz\ statistics as approximate corrections to the effective redshift distribution, and sample over them as nuisance parameters. The accuracy of such an approach should be validated against the use of the full corrections.

\section{Discussion}\label{sec:conclusions}

Image simulations have long been recognized as an essential tool for calibrating shear estimation pipelines, and have played a role in providing the shear calibration for all recent weak lensing cosmology analyses. While current methodologies like \mcal\ have made major strides in dealing with well-studied shear estimation biases like noise bias and model bias, we are now entering an era of weak lensing data that is both deep, resulting in significant blending of galaxy surface brightness profiles, and voluminous, thus statistically powerful enough to have very strict accuracy requirements. 

In this work we present realistic image simulations designed to accurately calibrate many of the complexities in analysing real imaging data. By using a combination of morphological information from HST and flux information from DES Deep Fields, we make use of an input galaxy sample with realistic joint distributions of flux (in multiple bands) and morphology. This, along with careful simulation of the observational characteristics of real DES Y3 data, gives us confidence that measurements of biases can be reliably applied as calibration corrections. This multi-band weak lensing calibration simulation allows us to perform the same photometric redshift inference as is performed on the real data, allowing reliable and consistent corrections for each photometric redshift bin. 

While we believe these simulations are sufficiently realistic given the requirements of the DES Year 3 cosmology analyses, they do of course involve approximations and shortcuts that may not be sufficiently accurate for future analyses. We do not simulate the estimation of the WCS solution, PSF, background levels or noise levels in our simulations, under the assumption that any errors in our estimation of these in the real data are negligible. These assumptions should be carefully re-examined for more precise upcoming weak lensing datasets.

In addition, we simulated galaxies with random sky positions, rather than a realistic level of clustering. While our re-weighting procedure in \sect{sec:m_popreweight} can increase the number of pairs of close detected objects, it does not  explicitly match the number of groups of more than two close detected objects, however, at current number densities, we expect those to be much less frequent. The re-weighting also cannot impact the statistics of undetected objects, and so will not change the frequency of detected galaxies having close, undetected neighbours. This specific case is investigated in detail by \citet{martinet19}, who demonstrate that for Euclid-like data, the clustering of faint galaxies around the bright ones used for shape measurement can induce a multiplicative bias of $\sim-0.5\%$ (with respect to the case where the faint galaxies are unclustered). Since this is of the same order as our statistical uncertainties, it is worth careful consideration. 

We firstly note that the only conclusive way to judge the impact of the clustering of undetected galaxies is to simulate it. However, generating a weak lensing calibration image simulation with realistic clustering statistics, as well as the realistic distributions of measured properties demonstrated here (\sect{sec:simval}) is an ongoing challenge. Current cosmological simulations which include the former have not been demonstrated to produce the high level of agreement with real data at the image level that is required for weak lensing calibration. In the absence of such a simulation, we make some arguments that the size of the biases estimated by \citet{martinet19} are likely an upper limit on the size of biases likely to be generated by clustering of undetected objects in DES Year 3. 

Firstly, the DES Year 3 data is shallower than the Euclid-like simulations in \citet{martinet19}, hence in general blending-related biases are lessened. Secondly, the ``faint" galaxies in \citet{martinet19} are defined as having $\snr<10$. We include all objects detected by \sex\ (and with positive flux) in our re-weighting procedure, which therefore only misses out on the $\snr<<10$ objects that are undetected by \sex. Thirdly, we are using \mcal, which has some robustness to blending - for very close pairs which are always detected as a single object, \mcal\ performs extremely well (as demonstrated in \citet{sheldon2020}), because it calibrates the response of the measurement to shear of all the light in the system. This is not the case for the shape measurement methods used in \citet{martinet19}, which we would expect to be more sensitive to ``model bias'' arising from the unusual apparent morphology of blended systems. Similarly, \textsc{metacalibration}'s insensitivity to noise bias means faint objects that are always undetected are unlikely to cause significant biases; their contribution behaves largely as extra noise. 

While these arguments are somewhat based on intuition rather than extensive simulation results, given constraints on time and computing resources, we think they justify our approach and priorities for the calibration of this intermediate dataset. Nonetheless, we recommended that \citet{y3-cosmicshear1} include a robustness test where an additional multiplicative bias with prior width $1\%$ was marginalized over in the cosmological inference from cosmic shear. This extra bias was assumed to be coherent across redshift bins, and represents a pessimistic interpretation of the potential bias arising from the effect explored by \citet{martinet19}. They find negligible change in the constraint on $S_8 = \sigma_8(\Omega_m/0.3)^{0.5}$, likely due to the dominance of other sources of systematic uncertainty.

Another potential limitation of our simulations is the limited size of our input galaxy catalog (see e.g. \citealt{Kannawadi15} for an investigation of the impact of limited input catalog size in weak lensing image simulations), which came from the COSMOS region, motivated by the availability of HST imaging which provides precise morphological information for our input galaxies. While our detailed comparisons of the simulation outputs and the real DES Year 3 data suggest that cosmic variance in the input catalog is not a huge effect (and this conclusion agrees with that of \citealt{kannawadi2019}), we would recommend investigations of the possibility of larger reliable input catalogs from deep ground-based imaging only (as measured by \citealt*{y3-deepfields} and used in injection simulations by \citealt{y3-balrog}). The planned LSST deep-drilling fields would be a valuable source of such data.

Conceptually, we have stressed the importance of characterizing biases due to blending via their impact on the effective redshift distribution. When blending becomes an important contributor to shear calibration biases, \photoz\ and shear inference can no longer be cleanly decoupled. Folding both \photoz\ and shear calibration biases into an effective redshift distribution for lensing, \neffz, is both a compact approach, and very general in that \neffz\ is the key observational quantity required to make theoretical predictions for most weak lensing signals of interest. The normalization of \neffz\ corresponds to the traditional mean multiplicative bias, $1+m$. 

Via simulations with redshift-dependent shear signals, we have provided a  methodology for directly estimating \neffz\ integrated over finite redshift intervals. The key point in this approach is that the value of the effective redshift distribution at a given redshift is defined by the response of the ensemble to shear at that redshift. Therefore only by simulating such a shear signal can the effective redshift distribution be estimated robustly. We believe that with the increased depth (and therefore increased blending) and stricter accuracy requirements of future surveys such as the Rubin Observatory and Euclid, such an approach will become essential i.e. \neffz\ is the quantity that must be accurately characterized for reliable cosmological inference, which is a requirement beyond the traditional approach of characterizing the multiplicative bias $m$.

In \sect{sec:y3nz} we propagated our simulation measurements through to corrections for the DES Y3 redshift distributions, producing from an input ensemble of redshift distributions, an output ensemble of effective redshift distributions, \neffz\, ready for use in the Year 3 weak lensing cosmology analyses. This output ensemble includes both statistical (due to finite simulation volume) and systematic (due to potential simulation inaccuracies) uncertainties. These \neffz\ should be used directly in the theoretical predictions for weak lensing statistics like the shear correlation functions and galaxy-galaxy lensing tangential shear, as the effective redshift distribution entering the  lensing kernel. Note that the (in general non-unity) normalization of the effective redshift distributions should be retained in the calculations, or applied to the resulting theory prediction if using a numerical code that normalizes the redshift distribution internally. 
This data will be made available along with the DES Year 3 shear catalogs, on publication of the key Year 3 cosmology analyses, 
as part of the DES Y3 coordinated release\footnote{\url{https://des.ncsa.illinois.edu/releases}}.

\section*{Acknowledgements}

This paper has gone through internal review by the DES collaboration. We thank the internal reviewers for their very helpful suggestions for improving the paper.

Funding for the DES Projects has been provided by the U.S. Department of Energy, the U.S. National Science Foundation, the Ministry of Science and Education of Spain, 
the Science and Technology Facilities Council of the United Kingdom, the Higher Education Funding Council for England, the National Center for Supercomputing 
Applications at the University of Illinois at Urbana-Champaign, the Kavli Institute of Cosmological Physics at the University of Chicago, 
the Center for Cosmology and Astro-Particle Physics at the Ohio State University,
the Mitchell Institute for Fundamental Physics and Astronomy at Texas A\&M University, Financiadora de Estudos e Projetos, 
Funda{\c c}{\~a}o Carlos Chagas Filho de Amparo {\`a} Pesquisa do Estado do Rio de Janeiro, Conselho Nacional de Desenvolvimento Cient{\'i}fico e Tecnol{\'o}gico and 
the Minist{\'e}rio da Ci{\^e}ncia, Tecnologia e Inova{\c c}{\~a}o, the Deutsche Forschungsgemeinschaft and the Collaborating Institutions in the Dark Energy Survey. 

The Collaborating Institutions are Argonne National Laboratory, the University of California at Santa Cruz, the University of Cambridge, Centro de Investigaciones Energ{\'e}ticas, 
Medioambientales y Tecnol{\'o}gicas-Madrid, the University of Chicago, University College London, the DES-Brazil Consortium, the University of Edinburgh, 
the Eidgen{\"o}ssische Technische Hochschule (ETH) Z{\"u}rich, 
Fermi National Accelerator Laboratory, the University of Illinois at Urbana-Champaign, the Institut de Ci{\`e}ncies de l'Espai (IEEC/CSIC), 
the Institut de F{\'i}sica d'Altes Energies, Lawrence Berkeley National Laboratory, the Ludwig-Maximilians Universit{\"a}t M{\"u}nchen and the associated Excellence Cluster Universe, 
the University of Michigan, NFS's NOIRLab, the University of Nottingham, The Ohio State University, the University of Pennsylvania, the University of Portsmouth, 
SLAC National Accelerator Laboratory, Stanford University, the University of Sussex, Texas A\&M University, and the OzDES Membership Consortium.

Based in part on observations at Cerro Tololo Inter-American Observatory at NSF's NOIRLab (NOIRLab Prop. ID 2012B-0001; PI: J. Frieman), which is managed by the Association of Universities for Research in Astronomy (AURA) under a cooperative agreement with the National Science Foundation.

The DES data management system is supported by the National Science Foundation under Grant Numbers AST-1138766 and AST-1536171.
The DES participants from Spanish institutions are partially supported by MICINN under grants ESP2017-89838, PGC2018-094773, PGC2018-102021, SEV-2016-0588, SEV-2016-0597, and MDM-2015-0509, some of which include ERDF funds from the European Union. IFAE is partially funded by the CERCA program of the Generalitat de Catalunya.
Research leading to these results has received funding from the European Research
Council under the European Union's Seventh Framework Program (FP7/2007-2013) including ERC grant agreements 240672, 291329, and 306478.
We  acknowledge support from the Brazilian Instituto Nacional de Ci\^encia
e Tecnologia (INCT) do e-Universo (CNPq grant 465376/2014-2).

This manuscript has been authored by Fermi Research Alliance, LLC under Contract No. DE-AC02-07CH11359 with the U.S. Department of Energy, Office of Science, Office of High Energy Physics.

NM acknowledges support from a European Research Council (ERC) Starting Grant under the
European Union's Horizon 2020 research and innovation
programme (Grant agreement No. 851274).
MRB is supported by DOE grant DE-AC02-06CH11357. 
This work was supported by the Department of Energy, Laboratory Directed Research and Development program at SLAC National Accelerator Laboratory, under contract DE-AC02-76SF00515 and as part of the Panofsky Fellowship awarded to DG.
MJ is partially supported by the US Department of Energy grant DE-SC0007901 and funds from the University of Pennsylvania.
AC acknowledges support from NASA grant 15-WFIRST15-0008.
RR, IH and SB acknowledge support from the European Research Council in the form of a Consolidator Grant with number 681431. IH also acknowledges support from the Beecroft Trust.

We gratefully acknowledge the computing resources 
provided on Bebop, a high-performance computing cluster operated by the Laboratory Computing Resource 
Center at Argonne National Laboratory, and the RHIC Atlas Computing Facility, operated by
Brookhaven National Laboratory. 

This research used resources of the National Energy Research Scientific Computing Center (NERSC), a U.S. Department of Energy Office of Science User Facility operated under Contract No. DE-AC02-05CH11231.

\section{Data Availability}

All DES Year 3 cosmology catalogs, including calibrations, will be made available at \url{https://www.darkenergysurvey.org/the-des-project/data-access/} on publication of the key cosmological  results.




\bibliographystyle{mnras_2author}
\bibliography{refs}



\onecolumn 
\appendix

\section{Shear and Redshift Bias Parameters used Prior to Unblinding the Analysis}

A final technical improvement to the priors on $m$ and \dz\ was completed in parallel with the unblinding of the DES Y3 3x2pt cosmology analysis. These updates were not expected to impact the analysis in a significant way. The original chains at unblinding used a different (but very similar) set of $m$ and \dz\ priors. Gaussian priors on the per-bin $m$ were marginalized over with mean, $\mu$ and width, $\sigma$ for the four \photoz\ bins given by $(\mu,\sigma) = \left[(-0.01044,0.00576), (-0.01579, 0.00414), (-0.02489, 0.00532), (-0.03802, 0.00801)\right]$. The mean of the \neffz\ samples generated using our fiducial correction was used, with a Gaussian \dz\ prior estimated from their statistical scatter only, with width $\left[0.018, 0.013, 0.006, 0.013\right]$.

\citet{y3-3x2ptkp} describe the unblinding criteria and procedure in detail, and conclude that updating the priors had negligible impact on the final cosmological constraints, and would not have affected the unblinding criteria (and therefore the decision to unblind).


\section{Simulation Validation Tests}\label{app:simval}

One of the main concerns when working with 
the complex simulation suite presented in this work is that the simulations themselves could contain bugs which would cause non-zero
multiplicative or additive biases. To guard against this possibility, we tested the performance of our pipelines
and the simulations in configurations where previous studies have demonstrated sub-$0.1\%$ performance. 

In these simulations, all objects have exponential light profiles. We fix the objects to have the same 
magnitude in each band, but different simulations vary the magnitude overall. The galaxies have a half-light 
radius that is either fixed to 0.5$\arcsec$ or allowed to vary according to the input catalogs 
for the more realistic simulations, although in the latter case, we limit the half-light radius to at most 0.8$\arcsec$ in order to 
prevent adjacent objects from overlapping. We test both a Gaussian PSF with a FWHM of 0.9$\arcsec$ and 
the smoothed \texttt{Piff} PSF models. We also vary whether or not we apply the bad pixel masks from the data to the simulations 
(and then correct for them). See \citet{y3-gold} and \citet*{y3-shapecatalog} for more details on the bad pixel masks and how our 
measurement codes correct for them. Further, we vary whether or not we perform pixel-level outlier rejection on the individual 
postage stamps for each object. Finally, we also vary a detail of how \mcal\ handles the computation of 
the PSF it uses in its reconvolution step. These variants are denoted internally as \texttt{symmetrize} or 
\texttt{gauss-fit}. The simulation configurations we tested are listed in Table~\ref{tab:basic-sims}. 

The measurements for $m$ and $c$ from these simulations are in the last two columns of Table~\ref{tab:basic-sims}.
We find that in the simplest case (configuration 1), we recover the applied shear to better than a part in a 
thousand with no evidence of any additive bias. The same conclusion holds for configurations 2, 3, and 4a which 
vary the galaxy sizes, the PSF model, and the \mcal\ reconvolution PSF treatment. Comparing configurations 
4a and 4, we find that galaxy size variation coupled with the \texttt{symmetrize} \mcal\ PSF treatment
introduces a small additive error in $g2$, but one that is too small to impact
cosmology. Configuration 5, which introduces pixel masking effects relative to configuration 4, demonstrates that 
our masking corrections cause a small $\sim-0.2\%$ multiplicative bias. This effect is well below our requirements 
and is subdominant relative to detection and/or blending, as described above. Configurations 6, 7, 8, and 
9 use the pixel-level outlier rejection and employ a series of fainter objects from magnitude 17 to magnitude 20.5. 
We find that extremely bright objects exhibit some non-trivial multiplicative bias in this case. This effect was 
traced back to Poisson noise from the object's flux falsely triggering the pixel-level outlier rejection 
(which only considered the sky noise in its threshold). In the data, these bright objects are extremely rare. 
Further, this effect is correctly modeled in our simulations for any objects that do eventually pass the cuts 
we employ on the shear catalogs.

\begin{table*}
\begin{tabular}{ ccccccclc }
 \hline
 config. \# & gal. mag. & gal. size & PSF model & masking effects & outlier rej. & mcal PSF & m ($\times10^{-3}$) & c ($\times10^{-5}$)\\
 \hline
 1  & 17    & 0.5$\arcsec$ & Gaussian      & no  & no  & $\texttt{gauss-fit}$  & g1: $0.225\pm0.005$ & $0.1\pm0.1$ \\ 
 \multicolumn{7}{c}{}                                                          & g2: $0.440\pm0.004$ & $-0.1\pm0.1$ \\
 2  & 17    & variable     & Gaussian      & no  & no  & $\texttt{gauss-fit}$  & g1: $0.314\pm0.007$ & $-0.2\pm0.2$ \\ 
 \multicolumn{7}{c}{}                                                          & g2: $0.415\pm0.007$ & $-0.1\pm0.02$ \\
 3  & 17    & variable     & Gaussian      & no  & no  & $\texttt{symmetrize}$ & g1: $0.381\pm0.009$ & $0.1\pm0.2$ \\ 
 \multicolumn{7}{c}{}                                                          & g2: $0.383\pm0.009$ & $-0.4\pm0.5$ \\
 4a & 17    & variable     & \texttt{Piff} & no  & no  & $\texttt{gauss-fit}$  & g1: $0.219\pm0.025$ & $-0.1\pm0.3$ \\ 
 \multicolumn{7}{c}{}                                                          & g2: $0.294\pm0.127$ & $0.1\pm0.3$ \\
 4  & 17    & variable     & \texttt{Piff} & no  & no  & $\texttt{symmetrize}$ & g1: $0.405\pm0.075$ & $0.2\pm0.3$ \\ 
 \multicolumn{7}{c}{}                                                          & g2: $0.414\pm0.014$ & $1.4\pm0.4$ \\
 5  & 17    & variable     & \texttt{Piff} & yes & no  & $\texttt{symmetrize}$ & g1: $-1.95\pm0.30$  & $0.6\pm0.4$ \\ 
 \multicolumn{7}{c}{}                                                          & g2: $-1.94\pm0.29$  & $6.8\pm1.0$ \\
 6  & 18    & variable     & \texttt{Piff} & no  & yes & $\texttt{symmetrize}$ & g1: $8.56\pm0.16$   & $0.0\pm0.9$ \\ 
 \multicolumn{7}{c}{}                                                          & g2: $8.70\pm0.13$   & $-9.5\pm1.0$ \\
 7  & 18.75 & variable     & \texttt{Piff} & no  & yes & $\texttt{symmetrize}$ & g1: $3.79\pm0.33$   & $-2.0\pm2.3$ \\ 
 \multicolumn{7}{c}{}                                                          & g2: $3.78\pm0.13$   & $-2.3\pm1.9$ \\
 8  & 19.5  & variable     & \texttt{Piff} & no  & yes & $\texttt{symmetrize}$ & g1: $2.44\pm0.16$   & $-2.8\pm2.4$ \\ 
 \multicolumn{7}{c}{}                                                          & g2: $2.50\pm0.11$   & $-1.0\pm2.6$ \\
 9  & 20.5  & variable     & \texttt{Piff} & no  & yes & $\texttt{symmetrize}$ & g1: $0.89\pm0.29$   & $-5.9\pm6.1$ \\ 
 \multicolumn{7}{c}{}                                                          & g2: $0.56\pm0.26$   & $0.1\pm6.6$ \\
 \hline
\end{tabular}
\caption{
  \label{tab:basic-sims} Validation simulation configurations and results. All simulations place the 
  galaxies on a grid and use ``true detection'' where the shear measurement code is fed the true position 
  of the galaxy in the image. We then run variations with and without 
  bad pixel masking and corrections ("masking effects"), pixel-level outlier rejection ("outlier rej."), 
  the galaxy and PSF models ("gal. mag", "gal. size", and "PSF model"), and the \mcal\ treatment of 
  the PSF model ("mcal PSF"). The last two columns list the multiplicative and additive biases 
  we find in these simulation setups. See the text in Appendix~\ref{app:simval} for details.
}
\end{table*}

\section{\texttt{Piff} PSF Model Smoothing for Simulations} \label{app:piffmod}

The simulations in this work use the observed \texttt{Piff} PSF models. We found a bug in the process of doing the DES Y3 analysis, 
detailed in \citet[\S3.2]{y3-piff}, which causes some images of the \texttt{Piff} PSF model to have what is effectively extra background noise.
In tests of the simulations, we found that when using the model images as the true PSF for the simulation, sometimes this extra noise caused 
shear recovery biases with \mcal\ even in cases where that method is known to work at extremely high precision. In order to 
avoid any potential problems in the simulations presented here, we applied the following smoothing procedure to the \texttt{Piff} PSF 
models. For each single-epoch image, we fit the rendered image of the \texttt{Piff} model with an \texttt{ngmix} \texttt{turb} model
(which is an approximation to a Kolmogorov-like PSF for a long exposure taken through atmospheric turbulence). This fit is done over a 
grid of locations on the CCD and the parameters of it are interpolated. We then use these interpolated fit parameters to draw the image of the PSF 
model at any location on the CCD. We include a small correction for the size of pixel in the fit parameters in order to more closely match
the output PSF sizes when the PSF is drawn. 

\begin{figure*}
    \centering
    \includegraphics[width=\columnwidth]{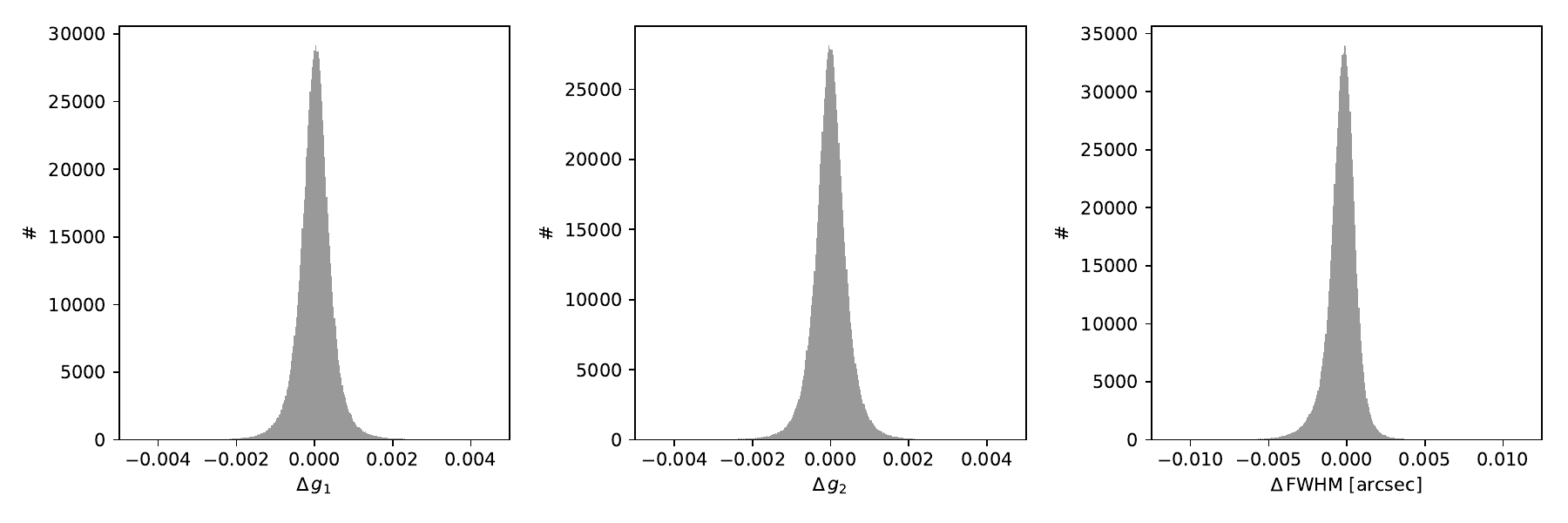}
    \caption{Performance of the smoothed \texttt{Piff} PSF models. Each panel from left to right shows a histogram of the differences 
    in the shape ($g_1$ or $g_2$) or FWHM of the smoothed PSF model versus the underlying \texttt{Piff} PSF model.}
    \label{fig:piff_psf}
\end{figure*}

Figure~\ref{fig:piff_psf} shows the differences in the shapes and FWHM for the raw \texttt{Piff} models and the rendered smooth models. 
To make these histograms, we selected $10^{4}$ CCDs at random, fit our smoothed, interpolated models, and then drew images of both the smooth model
and the interpolated one at 10 random points per CCD. We find an extremely small negative bias in the FWHM, but otherwise the models provide 
a good approximation to the true PSF size and shape variations.

\section{The Effect of Smooth Response Weighting on Redshift Distributions} \label{app:smoothR}
As previously noted in Section \ref{sec:ngamma}, the redshift distributions are weighted by the \texttt{METACALIBRATION} response. It was found in general that the noise in response measurement could result in unrealistic redshift distributions, where negatively measured responses could lead to an unphysical negative $N(z)$ at several points. Noise in response measurement then, tracks directly into N(z) scatter. 

\begin{figure*}
    \centering
    \includegraphics[width=0.5\linewidth]{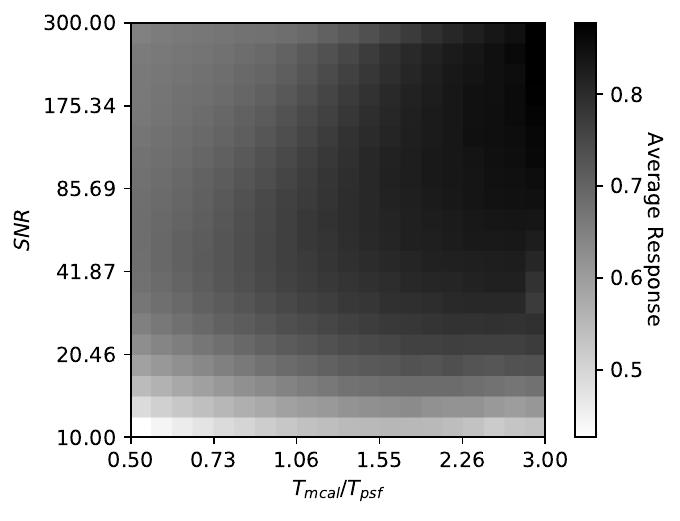}
    \caption{The smooth response grid generated from the DES Y3 catalog shows correlation between mean response and size ratio, and mean response and signal to noise. This 20 x 20 log-spaced grid supplied the smoothed weighting scheme evaluated in App. \ref{app:smoothR}. }
    \label{fig:rgrid}
\end{figure*}

The DES Year 3 analysis resolves this problem by smoothing  the same two sets of weights (shear and response) over a log grid in signal to noise \texttt{mcal\_s2n\_r} and the size ratio with respect to the point spread function, $T_\mathrm{mcal}/T_\mathrm{psf}$. One such grid, generated from the mesh average in the data is plotted in Fig. \ref{fig:rgrid}. For each detection in the image simulations, we assign it to a cell on this grid of SNR and size ratio, and take the mean response of all detections in that cell. This mean response (or \emph{smooth response}) is then used when weighting the redshift distribution. 

For this weighting scheme to be valid, we tested the shift in mean redshift for each tomographic bin in the image simulations from the reweighting, as this data set has a very well understood input redshift distribution. The shift applied as a result of smoothing in each bin was on average 0.0025 in mean redshift, with the first bin being the most affected with a shift of 0.004, and the third being the least affected at 0.0002. Response weighting as a whole shifts the mean redshift in each bin by an order of magnitude larger, so the majority of the effect is preserved. In conclusion, given the reduction in $N(z)$ scatter achieved with a smooth response weight and the small amplitude of bias ($\sim 10^{-3}$), this approximation can be made when evaluating the Year 3 data.

\section{Affiliations}
\input{affiliations.tex}


\bsp	
\label{lastpage}
\end{document}

%% file: authors.tex
\author[DES Collaboration]{
\parbox{\textwidth}{
\Large
N.~MacCrann,$^{1,\dag}$
M.~R.~Becker,$^{2}$
J.~McCullough,$^{3,4,5}$
A.~Amon,$^{3,4,5}$
D.~Gruen,$^{4,3,5}$
M.~Jarvis,$^{6}$
A.~Choi,$^{7}$
M.~A.~Troxel,$^{8}$
E.~Sheldon,$^{9}$
B.~Yanny,$^{10}$
K.~Herner,$^{10}$
S.~Dodelson,$^{11}$
J.~Zuntz,$^{12}$
K.~Eckert,$^{6}$
R.~P.~Rollins,$^{13}$
T.~N.~Varga,$^{14,15}$
G.~M.~Bernstein,$^{6}$
R.~A.~Gruendl,$^{16,17}$
I.~Harrison,$^{13,18}$
W.~G.~Hartley,$^{19}$
I.~Sevilla-Noarbe,$^{20}$
A.~Pieres,$^{21,22}$
S.~L.~Bridle,$^{13}$
J.~Myles,$^{4}$
A.~Alarcon,$^{2}$
S.~Everett,$^{23}$
C.~S{\'a}nchez,$^{6}$
E.~M.~Huff,$^{24}$
F.~Tarsitano,$^{25}$
M.~Gatti,$^{26}$
L.~F.~Secco,$^{6}$
T.~M.~C.~Abbott,$^{27}$
M.~Aguena,$^{28,21}$
S.~Allam,$^{10}$
J.~Annis,$^{10}$
D.~Bacon,$^{29}$
E.~Bertin,$^{30,31}$
D.~Brooks,$^{32}$
D.~L.~Burke,$^{3,5}$
A.~Carnero~Rosell,$^{33,34}$
M.~Carrasco~Kind,$^{16,17}$
J.~Carretero,$^{26}$
M.~Costanzi,$^{35,36}$
M.~Crocce,$^{37,38}$
M.~E.~S.~Pereira,$^{39}$
J.~De~Vicente,$^{20}$
S.~Desai,$^{40}$
H.~T.~Diehl,$^{10}$
J.~P.~Dietrich,$^{41}$
P.~Doel,$^{32}$
T.~F.~Eifler,$^{42,24}$
I.~Ferrero,$^{43}$
A.~Fert\'e,$^{24}$
B.~Flaugher,$^{10}$
P.~Fosalba,$^{37,38}$
J.~Frieman,$^{10,44}$
J.~Garc\'ia-Bellido,$^{45}$
E.~Gaztanaga,$^{37,38}$
D.~W.~Gerdes,$^{46,39}$
T.~Giannantonio,$^{47,48}$
J.~Gschwend,$^{21,22}$
G.~Gutierrez,$^{10}$
S.~R.~Hinton,$^{49}$
D.~L.~Hollowood,$^{23}$
K.~Honscheid,$^{7,50}$
D.~J.~James,$^{51}$
O.~Lahav,$^{32}$
M.~Lima,$^{28,21}$
M.~A.~G.~Maia,$^{21,22}$
M.~March,$^{6}$
J.~L.~Marshall,$^{52}$
P.~Martini,$^{7,53,54}$
P.~Melchior,$^{55}$
F.~Menanteau,$^{16,17}$
R.~Miquel,$^{56,26}$
J.~J.~Mohr,$^{41,14}$
R.~Morgan,$^{57}$
J.~Muir,$^{3}$
R.~L.~C.~Ogando,$^{21,22}$
A.~Palmese,$^{10,44}$
F.~Paz-Chinch\'{o}n,$^{47,17}$
A.~A.~Plazas,$^{55}$
M.~Rodriguez-Monroy,$^{20}$
A.~Roodman,$^{3,5}$
S.~Samuroff,$^{11}$
E.~Sanchez,$^{20}$
V.~Scarpine,$^{10}$
S.~Serrano,$^{37,38}$
M.~Smith,$^{58}$
M.~Soares-Santos,$^{39}$
E.~Suchyta,$^{59}$
M.~E.~C.~Swanson,$^{17}$
G.~Tarle,$^{39}$
D.~Thomas,$^{29}$
C.~To,$^{4,3,5}$
and R.D.~Wilkinson$^{60}$
\begin{center} (DES Collaboration) \end{center}
}
\vspace{0.4cm}
}

%% file: affiliations.tex
$^{1}$ Department of Applied Mathematics and Theoretical Physics, University of Cambridge, Cambridge CB3 0WA, UK\\
$^{2}$ Argonne National Laboratory, 9700 South Cass Avenue, Lemont, IL 60439, USA\\
$^{3}$ Kavli Institute for Particle Astrophysics \& Cosmology, P. O. Box 2450, Stanford University, Stanford, CA 94305, USA\\
$^{4}$ Department of Physics, Stanford University, 382 Via Pueblo Mall, Stanford, CA 94305, USA\\
$^{5}$ SLAC National Accelerator Laboratory, Menlo Park, CA 94025, USA\\
$^{6}$ Department of Physics and Astronomy, University of Pennsylvania, Philadelphia, PA 19104, USA\\
$^{7}$ Center for Cosmology and Astro-Particle Physics, The Ohio State University, Columbus, OH 43210, USA\\
$^{8}$ Department of Physics, Duke University Durham, NC 27708, USA\\
$^{9}$ Brookhaven National Laboratory, Bldg 510, Upton, NY 11973, USA\\
$^{10}$ Fermi National Accelerator Laboratory, P. O. Box 500, Batavia, IL 60510, USA\\
$^{11}$ Department of Physics, Carnegie Mellon University, Pittsburgh, Pennsylvania 15312, USA\\
$^{12}$ Institute for Astronomy, University of Edinburgh, Edinburgh EH9 3HJ, UK\\
$^{13}$ Jodrell Bank Centre for Astrophysics, School of Physics and Astronomy, University of Manchester, Oxford Road, Manchester, M13 9PL, UK\\
$^{14}$ Max Planck Institute for Extraterrestrial Physics, Giessenbachstrasse, 85748 Garching, Germany\\
$^{15}$ Universit\"ats-Sternwarte, Fakult\"at f\"ur Physik, Ludwig-Maximilians Universit\"at M\"unchen, Scheinerstr. 1, 81679 M\"unchen, Germany\\
$^{16}$ Department of Astronomy, University of Illinois at Urbana-Champaign, 1002 W. Green Street, Urbana, IL 61801, USA\\
$^{17}$ National Center for Supercomputing Applications, 1205 West Clark St., Urbana, IL 61801, USA\\
$^{18}$ Department of Physics, University of Oxford, Denys Wilkinson Building, Keble Road, Oxford OX1 3RH, UK\\
$^{19}$ D\'{e}partement de Physique Th\'{e}orique and Center for Astroparticle Physics, Universit\'{e} de Gen\`{e}ve, 24 quai Ernest Ansermet, CH-1211 Geneva, Switzerland\\
$^{20}$ Centro de Investigaciones Energ\'eticas, Medioambientales y Tecnol\'ogicas (CIEMAT), Madrid, Spain\\
$^{21}$ Laborat\'orio Interinstitucional de e-Astronomia - LIneA, Rua Gal. Jos\'e Cristino 77, Rio de Janeiro, RJ - 20921-400, Brazil\\
$^{22}$ Observat\'orio Nacional, Rua Gal. Jos\'e Cristino 77, Rio de Janeiro, RJ - 20921-400, Brazil\\
$^{23}$ Santa Cruz Institute for Particle Physics, Santa Cruz, CA 95064, USA\\
$^{24}$ Jet Propulsion Laboratory, California Institute of Technology, 4800 Oak Grove Dr., Pasadena, CA 91109, USA\\
$^{25}$ Department of Physics, ETH Zurich, Wolfgang-Pauli-Strasse 16, CH-8093 Zurich, Switzerland\\
$^{26}$ Institut de F\'{\i}sica d'Altes Energies (IFAE), The Barcelona Institute of Science and Technology, Campus UAB, 08193 Bellaterra (Barcelona) Spain\\
$^{27}$ Cerro Tololo Inter-American Observatory, NSF's National Optical-Infrared Astronomy Research Laboratory, Casilla 603, La Serena, Chile\\
$^{28}$ Departamento de F\'isica Matem\'atica, Instituto de F\'isica, Universidade de S\~ao Paulo, CP 66318, S\~ao Paulo, SP, 05314-970, Brazil\\
$^{29}$ Institute of Cosmology and Gravitation, University of Portsmouth, Portsmouth, PO1 3FX, UK\\
$^{30}$ CNRS, UMR 7095, Institut d'Astrophysique de Paris, F-75014, Paris, France\\
$^{31}$ Sorbonne Universit\'es, UPMC Univ Paris 06, UMR 7095, Institut d'Astrophysique de Paris, F-75014, Paris, France\\
$^{32}$ Department of Physics \& Astronomy, University College London, Gower Street, London, WC1E 6BT, UK\\
$^{33}$ Instituto de Astrofisica de Canarias, E-38205 La Laguna, Tenerife, Spain\\
$^{34}$ Universidad de La Laguna, Dpto. Astrofísica, E-38206 La Laguna, Tenerife, Spain\\
$^{35}$ INAF-Osservatorio Astronomico di Trieste, via G. B. Tiepolo 11, I-34143 Trieste, Italy\\
$^{36}$ Institute for Fundamental Physics of the Universe, Via Beirut 2, 34014 Trieste, Italy\\
$^{37}$ Institut d'Estudis Espacials de Catalunya (IEEC), 08034 Barcelona, Spain\\
$^{38}$ Institute of Space Sciences (ICE, CSIC),  Campus UAB, Carrer de Can Magrans, s/n,  08193 Barcelona, Spain\\
$^{39}$ Department of Physics, University of Michigan, Ann Arbor, MI 48109, USA\\
$^{40}$ Department of Physics, IIT Hyderabad, Kandi, Telangana 502285, India\\
$^{41}$ Faculty of Physics, Ludwig-Maximilians-Universit\"at, Scheinerstr. 1, 81679 Munich, Germany\\
$^{42}$ Department of Astronomy/Steward Observatory, University of Arizona, 933 North Cherry Avenue, Tucson, AZ 85721-0065, USA\\
$^{43}$ Institute of Theoretical Astrophysics, University of Oslo. P.O. Box 1029 Blindern, NO-0315 Oslo, Norway\\
$^{44}$ Kavli Institute for Cosmological Physics, University of Chicago, Chicago, IL 60637, USA\\
$^{45}$ Instituto de Fisica Teorica UAM/CSIC, Universidad Autonoma de Madrid, 28049 Madrid, Spain\\
$^{46}$ Department of Astronomy, University of Michigan, Ann Arbor, MI 48109, USA\\
$^{47}$ Institute of Astronomy, University of Cambridge, Madingley Road, Cambridge CB3 0HA, UK\\
$^{48}$ Kavli Institute for Cosmology, University of Cambridge, Madingley Road, Cambridge CB3 0HA, UK\\
$^{49}$ School of Mathematics and Physics, University of Queensland,  Brisbane, QLD 4072, Australia\\
$^{50}$ Department of Physics, The Ohio State University, Columbus, OH 43210, USA\\
$^{51}$ Center for Astrophysics $\vert$ Harvard \& Smithsonian, 60 Garden Street, Cambridge, MA 02138, USA\\
$^{52}$ George P. and Cynthia Woods Mitchell Institute for Fundamental Physics and Astronomy, and Department of Physics and Astronomy, Texas A\&M University, College Station, TX 77843,  USA\\
$^{53}$ Department of Astronomy, The Ohio State University, Columbus, OH 43210, USA\\
$^{54}$ Radcliffe Institute for Advanced Study, Harvard University, Cambridge, MA 02138\\
$^{55}$ Department of Astrophysical Sciences, Princeton University, Peyton Hall, Princeton, NJ 08544, USA\\
$^{56}$ Instituci\'o Catalana de Recerca i Estudis Avan\c{c}ats, E-08010 Barcelona, Spain\\
$^{57}$ Physics Department, 2320 Chamberlin Hall, University of Wisconsin-Madison, 1150 University Avenue Madison, WI  53706-1390\\
$^{58}$ School of Physics and Astronomy, University of Southampton,  Southampton, SO17 1BJ, UK\\
$^{59}$ Computer Science and Mathematics Division, Oak Ridge National Laboratory, Oak Ridge, TN 37831\\
$^{60}$ Department of Physics and Astronomy, Pevensey Building, University of Sussex, Brighton, BN1 9QH, UK\\